\documentclass[lettersize,journal]{IEEEtran}
\usepackage{amsmath,amsfonts}
\usepackage{amssymb}
\usepackage{algorithmic}
\usepackage{algorithm}
\usepackage{array}
\usepackage[caption=false,font=normalsize,labelfont=sf,textfont=sf]{subfig}
\usepackage{textcomp}
\usepackage{stfloats}
\usepackage{url}
\usepackage{verbatim}
\usepackage{graphicx}
\usepackage{cite}
\usepackage{orcidlink}
 \usepackage{balance}
\usepackage{multirow}
\usepackage{multicol}
\newcolumntype{d}[1]{D..{#1}}

\usepackage{booktabs}
\usepackage{tcolorbox}
\usepackage{flushend}
\usepackage{makecell}
\usepackage{colortbl}
\usepackage{color}
\usepackage{tablefootnote}
\usepackage{threeparttable} 

\begin{document}
\onecolumn
{\noindent\Large \textbf{IEEE Copyright Notice}}

${}$

{\noindent\large \copyright 2023 IEEE. 
Personal use of this material is permitted. Permission from IEEE must be obtained for all other uses, in any current or future media, including reprinting/republishing this material for advertising or promotional purposes, creating new collective works, for resale or redistribution to servers or lists, or reuse of any copyrighted component of this work in other works.

${}$

\noindent
Accepted by IEEE/ACM Transactions on Audio Speech and Language Processing, VOL. 31, 2023

${}$

\noindent
DOI: 10.1109/TASLP.2022.3233236
}

\twocolumn

\title{The PartialSpoof Database and Countermeasures \\ for the Detection of Short {Fake Speech} \\ Segments  Embedded in an Utterance}

\author{Lin Zhang \orcidlink{0000-0001-7826-2850}, \IEEEmembership{Student Member, IEEE}, Xin Wang \orcidlink{0000-0001-8246-0606}, \IEEEmembership{Member, IEEE}, Erica Cooper \orcidlink{0000-0002-2978-2793}, \IEEEmembership{Member, IEEE}, \\ Nicholas Evans \orcidlink{0000-0002-8459-1041}, \IEEEmembership{Member, IEEE}, and Junichi Yamagishi \orcidlink{0000-0003-2752-3955}, \IEEEmembership{Senior Member, IEEE}
\thanks{Manuscript received April 8, 2022; revised October 31 and December 14, 2022; accepted December 18, 2022. Date of publication December 30, 2022; date of current version December 28, 2022.

We would like to thank Prof.\ Kiyoshi Honda and Dr.\ Jose Patino for their comments. This study is partially supported by the Japanese-French joint national VoicePersonae project supported by JST CREST (JPMJCR18A6, JPMJCR20D3), JPMJFS2136 and the ANR (ANR-18-JSTS-0001), MEXT KAKENHI Grants (21K17775, 21H04906, 21K11951, 18H04112), Japan, and Google AI for Japan program.

The associate editor coordinating the review of this manuscript and approving it for publication was Prof. Sadjadi Omid. 
\textit{(Corresponding author: Lin Zhang.)}}
\thanks{Lin Zhang, Xin Wang, Erica Cooper and Junichi Yamagishi are with the National Institute of Informatics, 2-1-2 Hitotsubashi Chiyoda-ku, Tokyo 101-8340, Japan (e-mail: \{zhanglin, wangxin, ecooper, jyamagis\}@nii.ac.jp).}
\thanks{Nicholas Evans is with EURECOM, Campus SophiaTech, 450 Route des Chappes, 06410 Biot, France (e-mail: evans@eurecom.fr).}}

\markboth{IEEE/ACM TRANSACTIONS ON AUDIO, SPEECH, AND LANGUAGE PROCESSING,~Vol.~31, 2023 (DOI: 10.1109/TASLP.2022.3233236)}%
{Shell \MakeLowercase{\textit{et al.}}: The PartialSpoof Database and Countermeasures for the Detection of Short Generated Audio Segments Embedded in a Speech Utterance}

\maketitle

\begin{abstract}
Automatic speaker verification is susceptible to various manipulations and spoofing, such as text-to-speech synthesis, voice conversion, replay, tampering, adversarial attacks, and so on. We consider a new spoofing scenario called \textit{``Partial Spoof''} (PS) in which synthesized or transformed {speech} segments are embedded into a bona fide utterance. While existing countermeasures (CMs) can detect fully spoofed utterances, there is a need for their adaptation or extension to the PS scenario.

We propose various improvements to construct a significantly more accurate CM that can detect and locate short-generated spoofed {speech} segments at finer temporal resolutions. First, we introduce newly developed self-supervised pre-trained models as enhanced feature extractors. Second, we extend our PartialSpoof database by adding segment labels for various temporal resolutions. Since the short spoofed {speech} segments to be embedded by attackers are of variable length, six different temporal resolutions are considered, ranging from as short as 20 ms to as large as 640 ms. Third, we propose a new CM that enables the simultaneous use of the segment-level labels at different temporal resolutions as well as utterance-level labels to execute utterance- and segment-level detection at the same time. We also show that the proposed CM is capable of detecting spoofing at the utterance level with low error rates in the PS scenario as well as in a related logical access (LA) scenario. The equal error rates of utterance-level detection on the PartialSpoof database and ASVspoof 2019 LA database were {0.77 and 0.90\%, respectively. }

\end{abstract}

\begin{IEEEkeywords}
Anti-spoofing, deepfake, PartialSpoof, self-supervised learning, spoof localization, countermeasure
\end{IEEEkeywords}

\section{Introduction}
\label{sec:intro}

\IEEEPARstart{S}{peech} technologies play a crucial role in many aspects of life, e.g., keyword spotting in smart home devices, speaker recognition in online banking, diarization of meeting recordings, and speech recognition for captioning news broadcasts. However, such technologies are also vulnerable to spoofing –– synthesized, transformed, or manipulated speech can fool machines and even humans. A number of initiatives and challenges such as ASVspoof \cite{Wu2014, Kinnunen2017, Nautsch2021spoof19, asvspoof2021} have hence been organized to encourage research in countermeasure (CM) solutions, which are needed to protect speech applications and human listeners from spoofing attacks. Several types of spoofing scenarios have been considered and explored, including logical access (LA), DeepFake (DF), and physical access (PA). The LA scenario is designed for text-to-speech (TTS) synthesis and voice conversion (VC) attacks with and without telephony codecs, reflecting a use case of authentication in call centers. The DF scenario is similar to the LA scenario, but takes into account audio under strong compression for media streaming without an authentication process. The PA scenario targets the development of CMs against replay attacks. 
 
In the LA and DF scenarios, entire audio signals are generated using TTS or VC algorithms. Missing in past work is the consideration of scenarios in which synthesized or transformed speech segments are embedded into a bona fide speech utterance such that only a fraction of an utterance is spoofed. There are many possible motives for attackers to take such an approach. To give a few examples, specific words or phrases may be replaced with different ones, and negation words, such as ``not'' may be generated using TTS or VC, and inserted into an original utterance to completely change the meaning of a given sentence. {If an attacker has an audio file containing a phrase such as ``Search Google" for a particular person, the attacker can replace the word ``Search" with the synthesized phrase ``OK" and attempt a presentation attack against a text-dependent automatic speaker verification (ASV) system running on the person's device.} {An attacker could also use segments of units smaller than words. If an attacker synthesizes certain vowels and replaces the original vowels with the synthesized ones, he or she can manipulate words such as ``bat," ``bet," ``bit," ``bot," and ``but."} {If the attacker has knowledge of phonology, he or she can use even smaller units and manipulate the consonants /b/, /g/, and /d/ by synthesizing and replacing only the transitional part of the second formant, which is an acoustic cue for the consonants.} With modern speech-synthesis technologies having the ability to produce high-quality speech resembling a given target speaker's voice, these types of partial audio manipulations are becoming more likely to occur. We call this new spoofing scenario \textit{``Partial Spoof''} (PS).

We believe that speech utterances containing such short generated spoofed audio segments will likely be difficult to detect using CMs trained for the LA or DF scenarios since they typically use aggregation operations over time, in which case the short segments will have little bearing on the final CM score. Therefore, new CMs for the PS scenario are needed.

How should CMs be implemented for the PS scenario? We propose that CMs for the PS scenario should have two functions. The first involves the simple utterance-level detection of utterances containing any short spoofed segments. This is similar to the standard \emph{spoof} vs.\ \emph{bona fide} classification in the LA and DF scenarios. The second function is to detect which segments in an utterance are spoofed. This is to show which specific parts of an audio sample may have been generated using TTS or VC and to improve the explainability of the CMs. We refer to the former as ``utterance-level detection'' and the latter as ``segment-level detection.'' {While earlier we described examples of spoofing at the phone and word levels, in practice, there is no restriction on the speech unit to be used by attackers. Segment-level detection should be carried out without prior knowledge of the length of the unit used by the attacker and needs to support variable-length segments.}

In our previous study \cite{Zhang2021PartialSpoof}, we constructed a speech database called ``PartialSpoof'' designed for the PS scenario and reported initial results for utterance- and segment-level detection. In this preliminary work, segment-level detection was applied in a straightforward manner with a fixed temporal resolution. Whether the detection is per-utterance or per-segment, the features needed for CMs are likely to be similar. Therefore, in our subsequent work \cite{zhang21partialspoof_mtl}, we constructed a CM that simultaneously executes detection at the utterance level and (fixed) segment level using multi-task learning. However, these two studies showed that there is much room for improvement.

We thus extend our previous studies to improve the utterance- and segment-level detection of CMs in the PS scenario. We use self-supervised pre-trained models based on wav2vec 2.0 \cite{BaevskiZMA20-w2v2} and WavLM \cite{chen2021wavlm} as enhanced feature extractors. We also extend the PartialSpoof database. Specifically, we add segment-level labels for various temporal resolutions, instead of only for a fixed temporal resolution. The spoofed audio segments to be embedded by attackers are of variable length; thus, by using these labels, CM models can be trained to execute detection at various temporal resolutions. In the extended PartialSpoof database, there are six different temporal resolutions for these segment-level labels, ranging from as fine as 20 ms to as coarse as 640 ms. We also propose a CM that enables the simultaneous use of the segment-level labels at different temporal resolutions as well as utterance-level labels to execute utterance- and segment-level detection at the same time.

The remainder of this paper is structured as follows. In Section \ref{sec:PS}, we explain the PS scenario in more detail and review relevant topics. In Section \ref{sec:database}, we explain our PartialSpoof database with new segment labels at various temporal resolutions. In Section \ref{sec:overview_CM}, we define the tasks and overview existing CMs for the PS scenario. Then, in Section \ref{sec:proposed-cm}, we introduce the proposed CM. We present experiments and results in Section \ref{sec:experiments}, and conclude the paper and discuss future work in Section \ref{sec:conclusion}.

\section{What is the PS scenario?}
\label{sec:PS}

\subsection{Spoofing in the PS scenario and related scenarios}

In the PS scenario, we consider manipulated audio in which generated audio segments are embedded in a bona fide speech utterance and vice versa. The speaker characteristics of the embedded segments are similar to those of the true speaker. However, due to differences in the performance of TTS or VC methods, there are differences in speaker similarity. Although the PS scenario resembles other scenarios in which spoofing and CMs have already been widely studied, there are some crucial differences. 

The conventional LA and DF scenarios assume that speech in a single utterance is generated using a single TTS or VC method. In the PS scenario, however, a single speech utterance may contain audio segments generated using more than one TTS or VC method, even if the entire utterance consists only of spoofed segments.

The PS scenario is also closely related to copy-moves and splicing, which are scenarios well-studied for tampering forgery \cite{Zakariah2018-forgery,bevinamarad2020audioforgery}. The copy-move method of audio forgery involves the copying and pasting of segments within the same bona fide audio sample. Splicing forgery involves assembling a speech utterance using spliced segments obtained from other audio recordings\footnote{In the TTS field, ``unit selection'' techniques \cite{campbell1996chatr, PaulTaylor} apply a similar splicing operation on the basis of dynamic time warping to reduce concatenation artifacts.}. Spoofed audio in the PS scenario is hence a special case of splicing forgery in which segments do not come from other bona fide audio recordings but are generated using TTS or VC. 

\subsection{New realistic threats}
\label{sec:asv_partialspoof}

Certain companies have started to develop technologies and services that enable users to modify specific segments of a speech recording using TTS or VC without affecting other segments, making the final utterance match as closely as possible to the original, e.g., \cite{Tan_2021_editspeech, morrison2021context, Descript}. Although such technologies and services are desirable for users who would like to manipulate their own speech without re-recording, they increase the possibility of misuse in the form of impersonation and fraud. They may also pose a threat to other speech-based applications such as text-dependent ASV. There is thus a need to develop new CMs that are capable of detecting partially-spoofed utterances. 

{\subsection{Databases for PS scenario}
During the same time as when we built the initial PartialSpoof database in 2021 \cite{Zhang2021PartialSpoof}, another database was also proposed for the PS scenario \cite{Yi2021halftruth} for which a single multi-speaker TTS system was used to replace a single word within an utterance. This later became a part of the Audio Deep synthesis Detection (ADD) challenge database \cite{Yi2022ADD}. In addition to the limited number of TTS systems, the database has predetermined the target for replacement as a word, which is a form of prior knowledge even on the detection side. In contrast, we consider a spoofing scenario based on variable-length spoofed segments generated using several multi-speaker TTS and VC systems.
}

\section{New PartialSpoof database}
\label{sec:database}

With the goal of stimulating research on CM models suitable for the PS scenario, an early version of the PartialSpoof\footnote{{https://zenodo.org/record/5766198}} database \cite{Zhang2021PartialSpoof} containing the spoofed speech needed for model training was built and made available to the research community in 2021. In the PS scenario, attackers may create partially-spoofed audio in various ways. Attackers may insert segments of spoofed speech generated by TTS or VC into a natural utterance to add new content or substitute parts of the original utterance with spoofed speech to replace content. Alternately, instead of starting with a natural utterance and inserting or replacing segments with spoofed speech, they may start with an utterance generated by TTS or VC and insert or replace segments with natural speech, for example, in the case in which the attackers only have a small amount of bona fide data. Thus, it is necessary for the database to contain spoofed speech with different proportions of generated audio segments within a single utterance. We call the ratio of the duration of generated audio segments in an utterance to the total duration of the utterance ``intra-speech generated segment ratio.'' An efficient way to achieve this is to prepare pairs of speech utterances consisting of one that is entirely generated using TTS or VC and one original bona fide speech utterance then randomly substitute short segments within the pair. This enables us to create a large number of spoofed audio files with different intra-speech generated segment ratios. This is a key concept of the PartialSpoof database and we describe its construction policies and procedures as well as newly generated segment-level labels for a new version of the PartialSpoof database. 

\subsection{Construction policies}

Since the methods used by attackers are unknown in practice, these will likely be different from the training data used to train a real system. Therefore, we should assume that TTS/VC methods for generating spoofed audio segments in the training set are mostly different from those in the evaluation set. To achieve this, each subset of the ASVspoof 2019 LA database \cite{Wang2020data} was used to construct each corresponding subset of the PartialSpoof database. It was also assumed that the attacker could use generated audio segments at a variety of acoustic units, not limited to linguistic units such as words. Hence, we used variable-length speech segments found by voice activity detection (VAD) regardless of the content of the speech. 

{Finally, we assumed that the attacker could carry out not advanced but basic digital signal processing (DSP) to reduce artifacts of the spoofed audio for the purpose of a reliable assessment of CMs. Since it is unrealistic to assume that advanced processing will be carried out, the DSP and VAD used by the attacker were assumed to be basic methods, and certain errors and artifacts were assumed to occur naturally.} 

\subsection{Construction procedure}
\label{sec:database-procedures}
\begin{figure}[!t]
\centering

\includegraphics[width=1.0\linewidth]{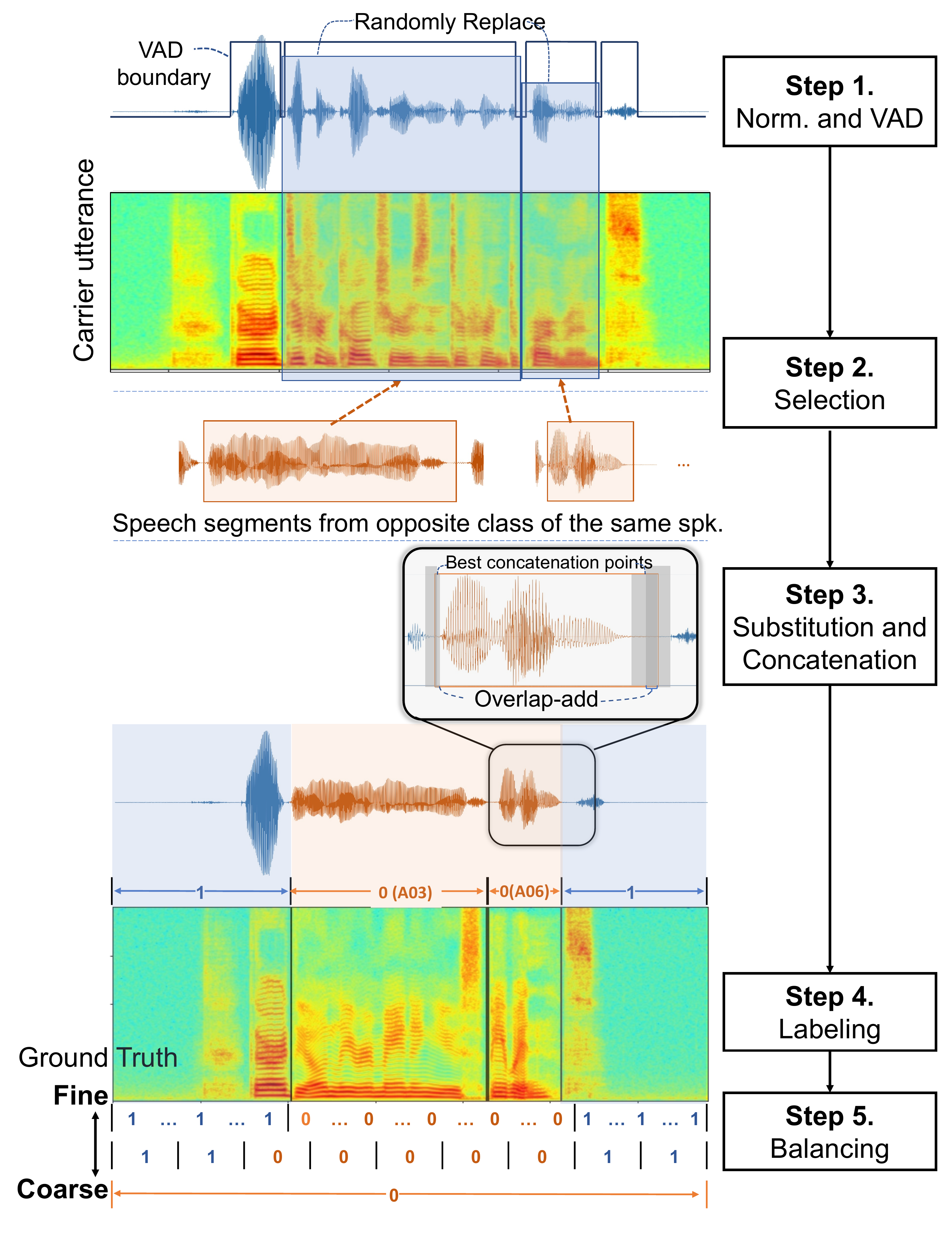}
\vspace*{-9mm}
\caption{Procedure to automatically construct our PartialSpoof database. (\textcolor{orange}{0} refers to \emph{spoof}, and \textcolor{blue}{1} refers to \emph{bona fide}.)}
\label{fig:database}
\vspace*{-5mm}
\end{figure}

The following is the procedure for constructing the PartialSpoof database automatically. As shown in Figure \ref{fig:database}, the entire processing pipeline involves five stages:

\vspace{1mm}
\noindent {\bf{Step 1. Normalization and VAD:}} When substituting segments into another speech utterance, the volume of the segments and utterance may be different. If there are significant volume mismatches, it becomes straightforward to detect the substituted segments, which would interfere with the CM's ability to learn fine-grained mismatches. Therefore, we first normalized the waveform amplitudes of the original and spoofed utterances contained within the ASVspoof 2019 LA database to $-26$\,dBov according to the ITU-T SV56 standard \cite{sv56}. 
Next, to select the variable-length candidate segments, three types of VAD \cite{povey2011kaldi, kinnunen2010overview, Lavechin-sad-dihard} were used, and a majority vote was taken. Segments detected as speech regions by two or more types of VAD within a human speech utterance were considered as candidate regions to be replaced with generated segments. Candidate segments within spoofed utterances were also found in the same manner.

\noindent {\bf{Step 2. Selection:}} 
The next step was to determine which segments found by the VAD should be replaced with which segments from the other class {of the same speaker.} {We considered both directions, that is, replacement of a randomly chosen segment from a bona fide utterance with a spoofed segment and substitution of a spoofed segment into a bona fide segment.}
The following conditions were used to select the segments: (1) segments are chosen {not from the same utterance but from different utterances of the same speaker;} (2) the same segment cannot be injected into an utterance more than once; (3) original and substitute segments must be of similar duration.

\noindent {\bf{Step 3. Substitution and concatenation:}} 
We then conducted substitution of the speech segments. To concatenate the segments without significant artifacts, we computed time-domain cross-correlation to find the best concatenation points. When searching for these points, parts of the silent regions around the corresponding speech were considered and the waveform overlap-add method was used.

\noindent {\bf{Step 4. Labeling:}} After the concatenation of the segments, each utterance was annotated with fine-grained segment labels at various temporal resolutions in addition to utterance-level labels. 
Speech frames generated by TTS/VC and those from the bona fide utterances were labeled as \emph{spoof} and \emph{bona fide}, respectively. Each segment or utterance that contains one or more generated frames was labeled as \emph{spoof}, otherwise \emph{bona fide}.

\noindent {\bf{Step 5. Post-processing:}} To balance the subsets, intra-speech generated segment ratios of the concatenated audio were quantized into ten levels, and then random sampling was done for each level. As a result, audio files with a small intra-speech generated segment ratio and those with a large intra-speech generated segment ratio were equally included in the database and in each subset.\footnote{The result of this process is that there is less mismatch between the subsets regarding the intra-speech generated segment ratio. It is also possible to design and adopt mismatched conditions; however, in this paper, we consider the matched condition to simplify our analysis.} 
The number of files that we randomly sampled was also the same as the \emph{spoof} class of the ASVspoof 2019 LA database.

\subsection{Database statistics}

\begin{table}[tb]
\caption{Details of trials in PartialSpoof dataset. Number of audio files, total duration, maximum number of unique TTS and VC methods within one utterance, and statistics of audio length for each subset are shown. }
\label{tab:data_stat}
\setlength{\tabcolsep}{3pt}
\begin{center}
\begin{tabular}{cccccccc}
\toprule
& \multirow{2}{*}{{Subset}} & \multirow{2}{*}{\shortstack{\# of \\ Samples}} & \multirow{2}{*}{\shortstack{Duration \\ (h)}} & \multirow{2}{*}{\shortstack{Max \# of \\ TTS/VC}} & \multicolumn{3}{c}{Audio length (s)} \\
\cline{6-8}
 & &  &  & & min & mean & max \\
\midrule
\multirow{3}{*}{\emph{bona fide}}  & {Train} & \phantom{0}2,580 & \phantom{0}2.43 & - & 1.36 & 3.39 & 11.13 \\
& {Dev.} & \phantom{0}2,548  & \phantom{0}2.48 & - & 1.28 & 3.51 & 11.39 \\
& {Eval.} & \phantom{0}7,355 & \phantom{0}6.94 & - & 0.90 & 3.40 & 13.01\\
\midrule
\multirow{3}{*}{\emph{spoof}} & {Train} & 22,800 & 21.82 & 6 & 0.60 & 3.45 & 21.02  \\
& {Dev.} & 22,296 & 21.86 & 6 & 0.62 & 3.53 & 15.34  \\
& {Eval.} & 63,882 & 60.74 & 9 & 0.48 & 3.42 & 18.20 \\
\bottomrule
\end{tabular}
\vspace{-5mm}
\end{center}
\end{table}

\begin{table}[tb]
\caption{{Number (in \textbf{thousands}) of samples in each temporal resolution.}}
\label{tab:data_realfake_tolnum}
\centering
\begin{tabular}{cccccccc}
\toprule
\multirow{2}{*}{{Subset}} & \multicolumn{6}{c}{{Temporal resolution} (ms)} \\
                   \cmidrule{2-7}
& {20} & {40} & {80} & {160} & {320} & {640} & {utt.} \\
\midrule
Train & \phantom{0}4,347  & 2,167 & 1,077 & \phantom{0}532   & 260 & 124 & 25 \\
Dev.  & \phantom{0}4,364  & 2,176 & 1,082 & \phantom{0}535   & 261 & 125 & 25 \\
Eval. & 12,129 & 6,047 & 3,006 & 1,485 & 725 & 346 & 71 \\
\bottomrule
\end{tabular}
\end{table}

\begin{table}[tb]
\caption{Percentages (\%) of \emph{spoof} class in each temporal resolution.}
\label{tab:data_realfake_ratio}
\centering
\begin{tabular}{cccccccc}
\toprule
\multirow{2}{*}{{Subset}}                   & \multicolumn{6}{c}{{Temporal resolution} (ms)} \\
                   \cmidrule{2-7}
& {20} & {40} & {80} & {160} & {320} & {640} & {utt.} \\
\midrule
{Train} & 43.79 & 45.10 & 47.76 & 53.00 & 61.51 & 73.77 & 89.83 \\
{Dev.}   & 42.96 & 44.30 & 47.02 & 52.31 & 60.86 & 73.41 & 89.74 \\
{Eval.}  & 38.46 & 39.81 & 42.60 & 48.03 & 57.52 & 71.19 & 89.68 \\
\bottomrule
\end{tabular}
\vspace{-5mm}
\end{table}

{The total number of samples for each temporal resolution is shown in Table \ref{tab:data_realfake_tolnum}. The sample size increases as resolution increases from utterance (utt.) to 20 ms.} 
The percentage of samples belonging to the \emph{spoof} class in each temporal resolution is shown in Table \ref{tab:data_realfake_ratio}. If any part of the generated waveform is contained in the target segment to be annotated, it is labeled as \emph{spoof}. We can see that the finer the temporal resolution of the segment, the less likely it is to contain generated audio signals; thus, the number of samples in the \emph{bona fide} class increases at finer-grained resolutions. As can be seen from the table, the data have a bias toward the \emph{spoof} class on a per utterance basis, but the extreme bias is eliminated on finer segment bases. In multi-task learning using labels at multiple temporal resolutions, such differences in bias may lead to more robust model learning.

\subsection{{Limitations}}

{In the PartialSpoof database, the variable-length speech segments found by the VADs are replaced without considering the meaning of sentences and words as well as the phonemes before and after the segments. Therefore, the speech in the database is not partially-spoofed speech that is intended to deceive listeners and deliver wrong linguistic messages to them but an approximation of it. On the other hand, since partially-spoofed segments of the PartialSpoof database are of variable length, CMs built on this database can execute segment-level detection at various temporal resolutions, and their accuracy can be evaluated.} 

\section{Overview of countermeasures for PS scenario}
\label{sec:overview_CM}

In this section we switch to the defender's side and explain the tasks of CMs and existing CM architectures for the PS scenario. We use the notations listed in Table~\ref{tab:sym} to facilitate the explanation.

\subsection{Task definition}
\label{sec:overview-task}

A CM for the PS scenario is required to conduct utterance- and segment-level detection on an input waveform $\boldsymbol{x}_{1:T}$. The two tasks can be defined more formally as follows.
\begin{tcolorbox}
\begin{itemize}
\item \textbf{Utterance-level detection:} learn a function $f_\theta$ to convert $\boldsymbol{x}_{1:T}$ to an utterance-level score $s^u$
\end{itemize}
\begin{equation}
f_\theta: \mathbb{R}^{T\times 1} \rightarrow \mathbb{R};\phantom{x}\boldsymbol{x}_{1:T} \mapsto s^u,
\end{equation}
\begin{itemize}
\item \textbf{Segment-level detection:} learn a function $f_\rho$ to convert $\boldsymbol{x}_{1:T}$ to segment-level scores $\boldsymbol{s}_{1:M}$
\end{itemize}
\begin{equation}
f_\rho: \mathbb{R}^{T\times 1} \rightarrow \mathbb{R}^{M\times 1};\phantom{x}\boldsymbol{x}_{1:T}\mapsto \boldsymbol{s}_{1:M}.
\end{equation}
\end{tcolorbox}
\noindent
Depending on the CM, $f_\theta$ and $f_\rho$ may share common architecture components and parameters since both tasks may be tackled by networks which operate upon similar hidden features.

In real applications, a CM needs to assign a \emph{bona fide} or \emph{spoof} label to the trial by comparing the $s^u$ with an application-dependent threshold. It may also classify each segment on the basis of $s_m$, $\forall{m}\in\{1, \cdots, M\}$. In this study, we follow the conventions in the research community and do not require the CM to produce hard decisions. Given $s^u$ and $\boldsymbol{s}_{1:M}$ computed from test set trials, we calculate EERs to measure the utterance- and segment-level detection performance of the CM. 

\begin{table}[t]
\caption{List of math notations used in this paper.}
\begin{center}
\setlength{\tabcolsep}{1pt}
\begin{tabular}{lp{5cm}}
\toprule
$x_t\in\mathbb{R}$ & a waveform sampling point at time $t$ \\
$\boldsymbol{x}_{1:T} = (x_1, x_2, \cdots, x_T)$ & a waveform with $T$ sampling points \\
$\boldsymbol{a}_n \in\mathbb{R}^{D_a}$ & an acoustic feature vector at the $n$-th frame\\
$\boldsymbol{a}_{1:N} = (\boldsymbol{a}_1,\boldsymbol{a}_2, \cdots, \boldsymbol{a}_N)$ & an acoustic feature sequence with $N$ frames \\
$\boldsymbol{h}_m \in\mathbb{R}^{D_h}$ & a hidden feature vector at the $m$-th segment \\
$\boldsymbol{h}_{1:M} = (\boldsymbol{h}_1, \boldsymbol{h}_2, \cdots, \boldsymbol{h}_M)$ & a hidden feature sequence with $M$ segments \\
$y^u\in\{\emph{bona fide}, \emph{spoof}\}$ & an utterance-level label \\
$\boldsymbol{e}^{u}\in\mathbb{R}^{D_e}$ & an utterance-level embedding vector \\
$s^u\in\mathbb{R}$ & an utterance-level CM score \\
$\boldsymbol{y}_{1:M}=(y_1, y_2, \cdots, y_M)$ & segment-level labels where $y_m$ is the label for the $m$-th segment\\
$\boldsymbol{e}_{1:M}=(\boldsymbol{e}_1, \boldsymbol{e}_2, \cdots, \boldsymbol{e}_M)$ &  segment-level embeddings where  $\boldsymbol{e}_m$ is for the $m$-th segment\\  
 $\boldsymbol{s}_{1:M} = (s_1, s_2, \cdots, s_M)$ & segment-level CM scores where $s_m$ is the score for the $m$-th segment\\
\bottomrule
\end{tabular}
\end{center}
\label{tab:sym}
\vspace{-4mm}
\end{table}

\subsection{Existing CM architectures for PS scenario}
\label{sec:previous-cms}

The definition of the utterance-level detection task in the PS scenario is identical to that in the LA (also PA and DF) scenario. Segment-level detection is different and more challenging because the required output is a sequence of scores, where each score must be derived from segments of comparatively shorter duration. Accordingly, CMs that conduct only utterance-level detection for the LA or other scenarios cannot be applied directly to the PS scenario.

Previous studies modified the utterance-level CMs for other conventional scenarios so that the CMs can do segment-level detection as well \cite{Zhang2021PartialSpoof, zhang21partialspoof_mtl,Yi2021halftruth}.
One approach involves the addition of a sub-network for segment-level detection to a conventional utterance-level CM. 
Another approach involves the application of a non-trainable re-scoring step to derive segment-level scores $\boldsymbol{s}_{1:M}$. We describe these modifications after first introducing the utterance-level CM. 

\begin{figure*}[!t]
\centering
\includegraphics[trim=0 150 0 105, clip, width=0.99\linewidth]{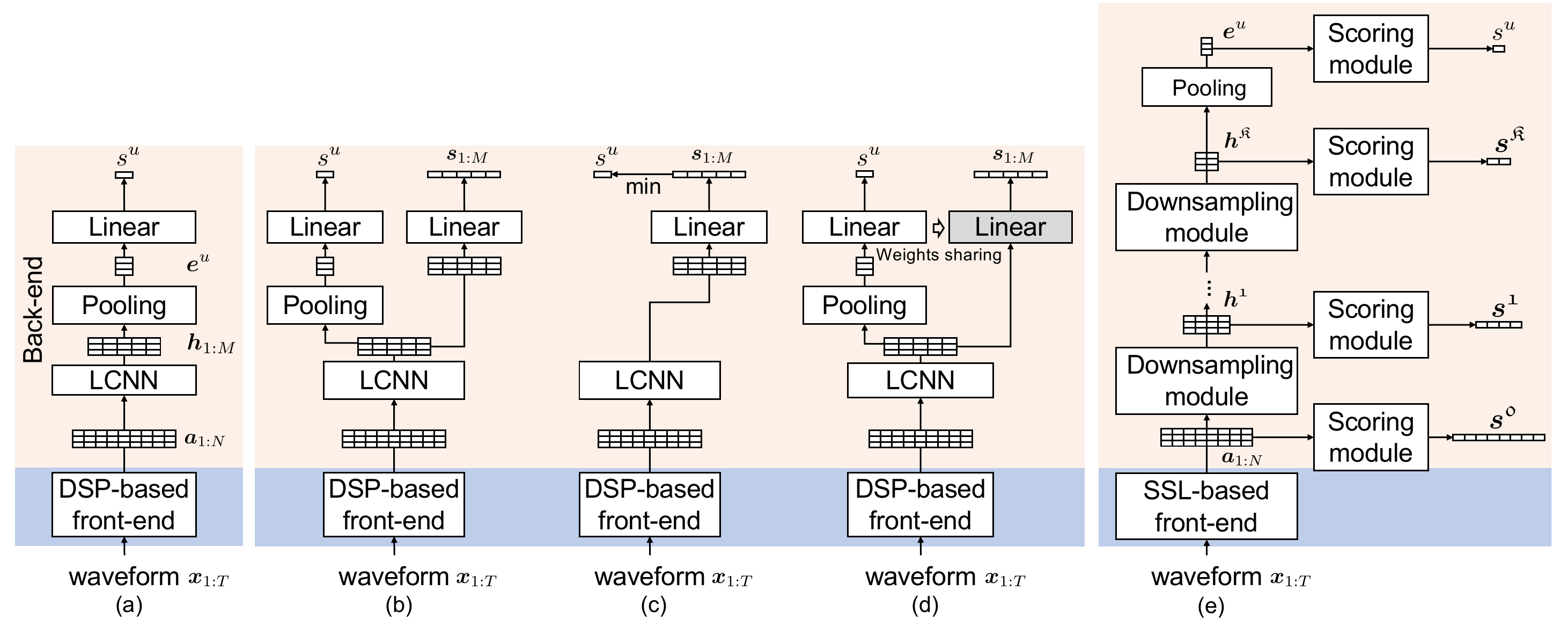} 
\vspace{-5mm}
\caption{Conventional utterance-level CM (a), existing CMs for PS scenario (b-d), and proposed CM for PS (e). Grey box in (d) indicates that linear layer is copied from the branch on its left side. {Conventional studies \cite{Zhang2021PartialSpoof,Yi2021halftruth,Yi2022ADD} and \cite{martin2022ADD1} used (a), \cite{zhang21partialspoof_mtl} and \cite{haibin2022partialQA} used (b), \cite{Zhang2021PartialSpoof, zhang21partialspoof_mtl} and \cite{Yi2021halftruth} used (c), and \cite{zhang21partialspoof_mtl} used (d).} Note that this figure illustrates the CMs during scoring. }
\label{fig:fully_CM}
\vspace{-3mm}
\end{figure*}

\subsubsection{\textbf{Conventional utterance-level CM}}
A conventional utterance-level CM consists of a front-end and back-end, as shown in Figure \ref{fig:fully_CM} (a). The front-end extracts an acoustic feature sequence $\boldsymbol{a}_{1:N}$ from the input $\boldsymbol{x}_{1:T}$, after which the back-end converts $\boldsymbol{a}_{1:N}$ into an utterance-level CM score $s^u$. The CM usually uses DSP algorithms in the front-end, and the conversion from $\boldsymbol{x}_{1:T}$ to $\boldsymbol{a}_{1:N}$ is hardwired by the DSP designer. In contrast, the back-end can be implemented using various machine learning models that can transform $\boldsymbol{a}_{1:N}$ into a scalar $s^u$.

We now explain a back-end based on a light convolutional neural network (LCNN) since it is related to this study and used in many CMs. It computes the utterance-level score by $\boldsymbol{a}_{1:N} \xrightarrow{\text{LCNN}} \boldsymbol{h}_{1:M} \xrightarrow{\text{Pooling}} \boldsymbol{e}^u \xrightarrow{\text{Linear}} s^u$, where $\boldsymbol{h}_{1:M}$ is a hidden feature sequence with length  $M \leq N$ determined by the LCNN stride, and $\boldsymbol{e}^u$ is an embedding vector of fixed dimension produced by global average pooling.
Finally, $s^u$ is computed from $\boldsymbol{e}^u$ using linear transformation. The back-end is typically trained by minimizing the standard cross-entropy (CE) or other advanced criteria given an utterance-level score $s^u$ and ground-truth label $y^u$ for a sufficiently large dataset. 

\subsubsection{\textbf{Adapting utterance-level CM for PS scenario}}
The conventional utterance-level CM has to be modified so that it can produce segment-level scores for the PS scenario.
Two studies proposed to revise the LCNN-based back-end \cite{Zhang2021PartialSpoof, Yi2021halftruth}. {As shown in Figure \ref{fig:fully_CM} (c), \cite{Zhang2021PartialSpoof} and \cite{Yi2021halftruth} remove pooling layer(s) from the utterance-level CM to derive $\boldsymbol{s}_{1:M}$ from $\boldsymbol{h}_{1:M}$.} While there are differences in the implementations, the common idea is to add a sub-network to transform hidden features into segment-level scores $\boldsymbol{h}_{1:M} \xrightarrow{\text{Linear}} \boldsymbol{s}_{1:M}$. The revised CM {may have} a bipartite structure as shown in Figure \ref{fig:fully_CM}(b). Note that the linear transformation is applied independently to each segment, i.e., $s_m = \text{Linear}(\boldsymbol{h}_{m}), \forall{m\in\{1, \cdots, M\}}$.

Segment scores $\boldsymbol{s}_{1:M}$ are deterministically aligned with the input $\boldsymbol{a}_{1:N}$. Specifically, if the LCNN uses a stride equal to 1 for all layers, $M$ will be equal to $N$, and each $s_m$ is aligned with $\boldsymbol{a}_{n=m}$\footnote{Strictly speaking, $s_m$ is computed given the acoustic features within the receptive field of the CNN, and the receptive field is centered around $\boldsymbol{a}_{n=m}$ and covers adjacent frames. The size of the receptive field is determined by the convolution kernel and dilation size. }. For strides greater than 1, the score-sequence length $M$ becomes shorter than $N$, and the $s_m$ is aligned with the $(\frac{N}{M}m)$-th input frame\footnote{For example, if the CNN consists of $k$ layers each with a stride equal to 2, $M$ will be equal to $N/2^k$, and $s_m$ will be aligned with $a_{n}$, where $n=2^km$. }. No matter how the two sequences are aligned,  $s_m$ reflects the degree to which the $m$-th segment is \emph{bona fide}, and thus can be used for segment-level detection. Training can be conducted by minimizing the CE loss between $\boldsymbol{s}_{1:M}$ and segment labels $\boldsymbol{y}_{1:M}$ and the utterance-level CE loss between the $s^u$ and $y^u$. This method is referred to as multi-task learning in our previous study \cite{zhang21partialspoof_mtl}.

It is also possible to carry out both utterance-level and segment-level detection using only the segment-level detection branch, as shown in Figure~\ref{fig:fully_CM}(c). Here, the utterance-level score $s^u$ is derived from the segment-level scores $\boldsymbol{s}_{1:M}$, for example, in accordance with $s^u = \min_{m} {{s}_{m}}$ \cite{zhang21partialspoof_mtl}. An advantage of this approach is the need to compute only the segment-level CE loss during training.

\subsubsection{\textbf{Attaching re-scoring step for PS scenario}}
The second approach involves not the adaptation of the utterance-level CM but the addition of a non-trainable re-scoring step to derive segment-level scores $\boldsymbol{s}_{1:M}$ \cite{Zhang2021PartialSpoof}. For the LCNN-based utterance-level CM, the transformation $\boldsymbol{h}_{1:M} \xrightarrow{\text{Pooling}} \boldsymbol{e}^u \xrightarrow{\text{Linear}} s^u$ can be written as $s^u=\boldsymbol{w}e^u=\frac{1}{M}\sum_{m=1}^{M}\boldsymbol{w}\boldsymbol{h}_{m}$, where $\boldsymbol{w}$ is the parameter of the linear layer. We can interpret $\boldsymbol{w}\boldsymbol{h}_{m}$ as a score for the $m$-th segment. Therefore, given the transformation vector $\boldsymbol{w}$ for utterance-level scoring, the re-scoring step to compute the $\boldsymbol{s}_{1:M}$ can be defined as 
\begin{align}
\boldsymbol{h}_{1:M} \xrightarrow{\text{Linear}} \boldsymbol{s}_{1:M} &: \quad s_m = \boldsymbol{w}\boldsymbol{h}_m, \forall m\in[1, M].
\end{align}
This approach is illustrated in Figure~\ref{fig:fully_CM}(d).

An advantage of this approach is that the utterance-level CM can be used directly for the PS scenario without further training. 
It does not require segment-level labels.
However, its potential is limited because without re-training, there is no adaptation to the spoofed segments in the PS scenario \cite{Zhang2021PartialSpoof}. 
   
{\subsection{CMs built for ADD challenge 2022}
One track of the ADD challenge 2022 is utterance-level detection of partially-spoofed audio \cite{Yi2022ADD}. Most participants used a conventional utterance-level CM, as shown in Fig.\ \ref{fig:fully_CM} (a) (e.g. \cite{martin2022ADD1}) whereas one team used a similar multi-task structure, as shown in Fig.\ \ref{fig:fully_CM} (b) \cite{haibin2022partialQA}.\footnote{{Interested readers are encouraged to see the challenge website and its details at http://addchallenge.cn/add2022.}} 
}

\section{Proposed CM for PS scenario}
\label{sec:proposed-cm}

\subsection{Motivation}
\label{sec:overview_CM_remark}

The above CMs have achieved encouraging results in existing studies \cite{Zhang2021PartialSpoof, Yi2021halftruth, zhang21partialspoof_mtl}, but there is still room for improvement. For example, compared with a deterministic DSP-based front-end, a trainable data-driven front-end may extract more discriminative features. 
As for the back-end, the commonly used LCNN only learns to conduct segment-level detection at a fixed temporal resolution. For example, using the LCNN configuration in \cite{Zhang2021PartialSpoof}, the back-end transforms the input $\boldsymbol{a}_{1:N}$ into hidden features $\boldsymbol{h}_{1:M}$, where $M=N/16$, and segment scores $s_m$ are computed once for every 16 frames. So we can hypothesize that it would be better to use a more flexible DNN architecture that can leverage segment labels at different temporal resolutions during training 
and conduct segment-level detection accordingly. 
The hypothesis motivated us to propose a new CM that consists of a self-supervised learning (SSL)-based front-end and a back-end that supports multiple temporal resolutions. The new front-end is expected to extract more discriminative acoustic features in a data-driven manner, while the new back-end enables the CM to better identify spoofed segments with varied length. 
Figure~\ref{fig:fully_CM}(e) illustrates the architecture of our CM\footnote{ https://github.com/nii-yamagishilab/PartialSpoof}. 

\subsection{Front-end}
An SSL speech model is a DNN that processes a waveform using trainable non-linear transformations. 
Because they are trained using task-agnostic self-supervised criteria on speech data from various domains, pre-trained SSL models can extract robust and informative acoustic features for many down-stream tasks \cite{yang21c_interspeech_superb}. SSL-based front-ends have also been shown to improve CM performance for the LA scenario \cite{yang21c_interspeech_superb, wang2021ssl, JoelShorSSL2021}.

Inspired by the above studies, we use an SSL-based front-end to extract acoustic features $\boldsymbol{a}_{1:N}$ from an input $\boldsymbol{x}_{1:T}$. While there are many types of SSL models, we investigated wav2vec 2.0 \cite{BaevskiZMA20-w2v2} and HuBERT \cite{hsu2021hubert}. These SSL models consist of a CNN-based encoder and cascade of Transformer blocks \cite{vaswani2017attention}. Following previous studies \cite{yang21c_interspeech_superb}, we extract output features from all the Transformer blocks and use their trainable weighted sum as $\boldsymbol{a}_{1:N}$. During CM training, we fine-tune the pre-trained SSL front-end in conjunction with the back-end. 

Note that the number of acoustic feature frames $N$ is determined by the waveform length $T$ and CNN encoder configuration. For the pre-trained SSL models we tested, the relationship is $N = T/320$. This means that, given an input waveform with a sampling rate of 16 kHz, the extracted acoustic features $\boldsymbol{a}_{1:N}$ have a `frame shift' of 20 ms. 

\subsection{Back-end}
Given $\boldsymbol{a}_{1:N}$, our back-end computes $\boldsymbol{s}_{1:M}$ at multiple temporal resolutions. This enables the proposed CM to handle the challenging segment-level detection task in a more flexible manner: by learning multiple functions $\{f_\rho^{\mathfrak{0}}, f_\rho^{\mathfrak{1}}, \cdots, f_\rho^{\mathfrak{K}} \}$, where the function $f_\rho^{\mathfrak{k}}$ converts $\boldsymbol{x}_{1:T}$ to segment-level scores $\boldsymbol{s}_{1:M^\mathfrak{k}}^{\mathfrak{k}}$ at the $\mathfrak{k}$-th temporal resolution, $\forall{\mathfrak{k}}\in\{\mathfrak{0},\cdots, \mathfrak{K}\}$. Without loss of generality, we assume there are $\mathfrak{K}$ temporal resolutions in total. Lower indices indicate finer temporal resolutions. To simplify the notation, we drop the subscript in data sequences (e.g., $\boldsymbol{s}_{1:M^{\mathfrak{k}}}^{\mathfrak{k}}$ becomes $\boldsymbol{s}^{\mathfrak{k}}$) for the rest of the paper.

As illustrated in Figure~\ref{fig:fully_CM}(e), the back-end computes $\boldsymbol{s}_{1:M}$ at different temporal resolutions in a fine-to-coarse order. First, the acoustic features $\boldsymbol{a}_{1:N}$ are used to compute the score sequence $\boldsymbol{s}^{\mathfrak{0}}\in\mathbb{R}^{N}$ at the frame level, which is the finest temporal resolution supported by the proposed CM. The $n$-th score ${s}^{\mathfrak{0}}_{n}$ indicates the likelihood that the $n$-th frame is \emph{bona fide}. Next, $\boldsymbol{a}_{1:N}$ is down-sampled to $\boldsymbol{h}^{\mathfrak{1}}$ and used to compute  $\boldsymbol{s}^{\mathfrak{1}}$. This process is repeated to compute segment-level scores for each temporal resolution. 

For utterance-level detection, the hidden feature sequences $\boldsymbol{h}^{\mathfrak{K}}$ at the lowest temporal resolution are pooled into the utterance-level embedding vector $\boldsymbol{e}^u$ which is then transformed into the utterance-level score $s^u$. This scoring module has the same architecture as those for segment-level detection. Combined with the front-end, the proposed CM computes the scores for the PS scenario in the following order:

\begin{center}
\setlength{\tabcolsep}{3pt}
\begin{tabular}{rrrcccc}
\hline
Input: & \multicolumn{4}{l}{$\boldsymbol{x}_{1:T}$} \\
Output: & \multicolumn{4}{l}{$\{\boldsymbol{s}^{\mathfrak{0}}, \boldsymbol{s}^{\mathfrak{1}}, \cdots, \boldsymbol{s}^{\mathfrak{K}}, {s}^{u} \}$} \\ 
\hline
Front-end: & $\boldsymbol{x}_{1:T}$  & $\xrightarrow{\text{\phantom{xxx}SSL model }}$ &  $\boldsymbol{a}_{1:N}$, \\
 Temporal res. $\mathfrak{0}$: & $\boldsymbol{a}_{1:N}$ & \multicolumn{3}{r}{$\xrightarrow{\text{\phantom{xxxxxxxxxxxxxxxxxxxxxxxxxxx}Scoring}}$} & $\boldsymbol{s}^{\mathfrak{0}}$, \\
 Temporal res.  $\mathfrak{1}$: & $\boldsymbol{a}_{1:N}$ & $\xrightarrow{\text{Downsampling}}$ & $\boldsymbol{h}^{\mathfrak{1}} $ & $\xrightarrow{\text{Scoring}}$ & $\boldsymbol{s}^{\mathfrak{1}}$, \\
 Temporal res.  $\mathfrak{2}$: & $\boldsymbol{h}^\mathfrak{1}$ & $\xrightarrow{\text{Downsampling}}$ & $\boldsymbol{h}^{\mathfrak{2}} $ & $\xrightarrow{\text{Scoring}}$ & $\boldsymbol{s}^{\mathfrak{2}}$, \\
 & $\cdots$ \\
 Temporal res.  $\mathfrak{K}$: & $\boldsymbol{h}^\mathfrak{K-1}$ & $\xrightarrow{\text{Downsampling}}$ & $\boldsymbol{h}^{\mathfrak{K}} $ & $\xrightarrow{\text{Scoring}}$ & $\boldsymbol{s}^{\mathfrak{K}}$, \\
 Utterance-level: & $\boldsymbol{h}^\mathfrak{K}$ & $\xrightarrow{\phantom{xxxxxx}\text{Pooling}}$ & $\boldsymbol{e}^{u} $ & $\xrightarrow{\text{Scoring}}$ & ${s}^{u}$, \\
 \hline
\end{tabular}
\end{center}

The scoring modules at different temporal resolutions are independent but use the same architecture, and 
we compare a few candidate architectures in Section~\ref{sec:exp-compare-front-back}. The down-sampling modules are also independent but use the same architecture. They contain a max-pooling operator with a stride equal to 2, followed by a 1D convolution layer with a kernel size of 1 and equal number of output and input channels. Therefore, if the input to the back-end $\boldsymbol{a}_{1:N}$ is of dimension $\mathbb{R}^{N \times D}$, $\boldsymbol{h}^{\mathfrak{k}}$ at the $\mathfrak{k}$-th temporal resolution will be in $\mathbb{R}^{N/2^\mathfrak{k} \times D}$, and the corresponding score sequence $\boldsymbol{s}^{\mathfrak{k}}$ is of dimension $\mathbb{R}^{N/2^\mathfrak{k}\times 1}$. This means that the temporal resolution of the $\boldsymbol{s}_{1:M}$ is reduced by a factor of 2 after each down-sampling module. 

Since acoustic features are extracted every 20 ms (i.e., the `frame shift' of the SSL-based front-end), 
a frame-level score $s^\mathfrak{0}_n$ is produced every 20 ms. This is the finest supported temporal resolution. For the $\mathfrak{k}$-th temporal resolution, one segment score $s_m$ is produced every  $20 \times 2^\mathfrak{k}$ ms. 
We set $\mathfrak{K}=5$, and the indices $\mathfrak{k}=0$ to $\mathfrak{k}=5$ correspond to temporal resolutions of 20, 40, 80, 160, 320, and 640 ms, respectively. {This setting was determined empirically, taking into account the ground-truth labels provided in the extended PartialSpoof database. Although the temporal resolutions do not have explicit linguistic meanings, we hope that each time resolution may capture the following content:} 
\begin{itemize}
    \item {40 ms: consonants and parts of vowels,}
    \item {80 and 160 ms: consonants and vowels, and}
    \item {320 ms: monosyllabic words and parts of longer words.}
\end{itemize}

\subsection{Training strategies}

\label{sec:proposed-training-strategies}
Supposing that ground-truth labels are available for all temporal resolutions, the proposed CM can be trained using two types of strategies:

\subsubsection{{Multiple temporal resolutions}}

For the $\mathfrak{k}$-th temporal resolution, a loss $\mathcal{L}^{\mathfrak{k}}$ is computed given the score $\boldsymbol{s}^{\mathfrak{k}}$ and label $\boldsymbol{y}^{\mathfrak{k}}$. Similar to the CMs described in Section \ref{sec:overview_CM}, the loss function can be CE or other advanced metrics. The gross loss function can thus be written as $\mathcal{L} = \sum_{\mathfrak{k}=0}^{\mathfrak{K+1}}\mathcal{L}^{\mathfrak{k}}$ \footnote{{In practice, the CM can be trained initially for the finest temporal resolution before more coarse resolutions, or the other way around. Furthermore, we may either fix or continue to fine-tune the scoring modules of a previous temporal resolution when moving to the next one. However, no obvious significant difference was observed among the aforementioned methods in our preliminary experiments. Hence, in this study we trained the CM at all temporal resolutions simultaneously.}}, where we use $\mathcal{L}^{\mathfrak{K}+1}$ to denote the utterance-level loss.

\subsubsection{{Single temporal resolution}}
\label{sec:proposed-training-tailored}
{As described earlier, CM training using multiple temporal resolutions requires} ground-truth labels for all temporal resolutions. While such labels are provided with the PartialSpoof database, they might not be available for other databases. In such a case, the CM can still be learned without application of the scoring modules and loss functions at temporal resolutions that have no segment-level labels. A special case is the learning of a CM for a single temporal resolution. For example, if we only have segment-level labels at the $\mathfrak{k}$-th temporal resolution, training can be carried out using only the back-end layers required to compute $\boldsymbol{s}^\mathfrak{k}$, in which case the training loss is computed over $\boldsymbol{s}^\mathfrak{k}$ and $\boldsymbol{y}^{\mathfrak{k}}$. After training, the CM can only produce scores $\boldsymbol{s}^\mathfrak{k}$, but not scores for other segment or utterance levels.

\section{Experiments}
\label{sec:experiments}

We conducted experiments to test the proposed CM on the PartialSpoof database. We first conducted a pilot study to select the most suitable SSL-based front-end and back-end scoring module. We then conducted a comparative study to compare {the two} CM training strategies. {After that, we analyzed differences in the tendencies of the proposed CM's performance on the development and evaluation sets}. Furthermore, we conducted a cross-scenario experiment using the ASVspoof 2019 LA database \cite{Wang2020data}, the goal of which was to show that the proposed CM optimized for the PS scenario can also be applied to the LA scenario. After explaining the experimental setup and model configurations, we describe each experiment in detail. 

\subsection{Experimental setup and model configurations}
All CMs were trained by minimizing the so-called P2SGrad-based mean square error \cite{wang2021comparative}. This criterion was used because it was found to be more stable and slightly superior to the conventional CE used in our previous study related to the LA scenario \cite{wang2021comparative}. During training, we used the Adam optimizer with a default configuration ($\beta_1=0.9, \beta_2=0.999, \epsilon=10^{-8}$). The learning rate was initialized with $1 \times 10^{-5}$ and halved every 10 epochs. 

We did not use any data augmentation, voice activity detection, or feature normalization during training, nor did we trim the input trials. All experiments were repeated three times with different random seeds for CM initialization, except for the pre-trained SSL front-end. The averaged results of the three runs are reported. Each round of training was conducted using a single Nvidia Tesla A100 card.

\begin{table}[t]
\caption{SSL models mentioned in Section~\ref{sec:exp-compare-front-back}. }
\vspace{-5mm}
\begin{center}
\resizebox{\columnwidth}{!}{
\setlength{\tabcolsep}{2pt}
\begin{tabular}{llp{2.7cm}rr}
\toprule
 ID & Model type & Data for pre-training & \# of paras & Feat. dim. \\
\midrule
\texttt{w2v2-base} & Wav2vec 2.0 Base & Librispeech & 95.04 m & 768 \\
\texttt{w2v2-large} & Wav2vec 2.0 Large & CommonVoice, Switchboard \cite{godfrey1992switchboard}, Libri-Light \cite{kahn2020libri}, Fisher \cite{cieri-etal-2004-fisher} & 317.38 m & 1024 \\
\texttt{w2v2-xlsr} & Wav2vec 2.0 Large  &  LibriSpeech \cite{panayotov2015librispeech}, CommonVoice \cite{ardila-etal-2020-common}, BABEL \cite{harper2011iarpa}  & 317.38 m & 1024 \\
\texttt{wavlm-large} & HuBERT Large & Libri-Light \cite{kahn2020libri}, GigaSpeech \cite{chen21o_GigaSpeech}, VoxPopuli \cite{wang-etal-2021-voxpopuli} &  316.62 m & 1024 \\
\bottomrule
\end{tabular}
}
\end{center}
\label{tab:ssl}
\end{table}

\begin{table}[!t]
\centering
\begin{center}
\caption{Comparison of different front-ends and back-ends.}
\label{tab:comparison}
\begin{tabular}{ccc}
\toprule
{Front-end}  & {Back-end}  & \multicolumn{1}{c}{{EER (\%) on dev. set}} \\
\midrule
\multirow{5}*{\texttt{w2v2-base}} & 1 FC           & 2.89                                  \\
          & 1 BLSTM + 1 FC  & 2.84                                  \\
          & 2 BLSTM + 1 FC & 2.49                                  \\
          & 1 gMLP             & 4.28                                  \\
          & 5 gMLP             & \textbf{2.22}                                  \\
\midrule
\texttt{w2v2-base}   & \multirow{4}*{5 gMLP}      & 2.22   \\
          
\texttt{w2v2-large} &     & \textbf{0.35}                         \\
\texttt{w2v2-xlsr}                &              & 0.65                                  \\
\texttt{wavlm-large}         &             & 0.64    
          \\
\bottomrule
\end{tabular}
\end{center}
\vspace{-5mm}
\end{table}

\subsection{Comparing front- and back-ends for proposed CM}
\label{sec:exp-compare-front-back}
As explained in Section~\ref{sec:proposed-cm}, we used a pre-trained SSL model for the front-end and designed a powerful scoring module for the back-end. In this experiment, we compared several candidates for both. Performance was measured using the utterance-level detection EER on the development set of the PartialSpoof database. The CM was trained following the same conventional utterance-level training as in the LA scenario, that is, using the single temporal resolution training strategy with utterance-level labels only.

For the front-end, we tested the four pre-trained SSL models listed in Table~\ref{tab:ssl}; \texttt{w2v2-base} is based on the Wav2vec 2.0 Base model, while \texttt{w2v2-large} and \texttt{w2v2-xlsr} are based on the Wav2vec 2.0 Large model. The difference is that the Wav2vec 2.0 Base model has only 12 Transformer \cite{vaswani2017attention} blocks while Wav2vec 2.0 Large has 24. There are further differences in the dimension of the output features of the Transformer block, as shown in Table~\ref{tab:ssl}. The \texttt{w2v2-large} and \texttt{w2v2-xlsr} models, both based on Wav2vec 2.0 Large, 
differ in the data used for pre-training. These three SSL models were included in the experiment because they have been shown to give reliable performance for the LA scenario \cite{wang2021ssl}. The last candidate \texttt{wavlm-large} is based on HuBERT \cite{hsu2021hubert}. It was included because it was the top entry on the leaderboard of SUPERB\footnote{SUPERB: https://superbbenchmark.org/} when we started this study.

For the scoring module in the back-end, we compared the following architectures:
\begin{itemize}
    \item a single fully-connected (FC) layer;
    \item an FC layer after a bidirectional long short-term memory (BLSTM) \cite{hochreiter1997long} layer;
    \item an FC layer after two BLSTM layers;
    \item a single gated multilayer perceptron (gMLP) block \cite{Liu2021gmlp};
    \item five gMLP blocks.
\end{itemize}
A gMLP block is similar to basic Multilayer Perceptrons (MLPs) with a gating unit. It was found to be simple and powerful compared with other alternatives.

We compared each of the above components in terms of the EER on the development set of the PartialSpoof database. To reduce the time cost, we used \texttt{w2v2-base} as the front-end when comparing the back-end scoring modules. The EERs listed in Table~\ref{tab:comparison} show that using five gMLP blocks gives the best performance. We then compared the SSL models while using the architecture of five gMLP blocks in the back-end scoring module. The results indicate that the three Large SSL models outperformed \texttt{w2v2-base}, and that \texttt{w2v2-large} performed best. Given these results, all following experiments were conducted using \texttt{w2v2-large} and five gMLP layers.

\subsection{Comparing training strategies on the proposed CM}
\label{sec:exp-single-multi-reso}

\begin{figure*}[!t]
\centering
\includegraphics[width=1.0\linewidth]{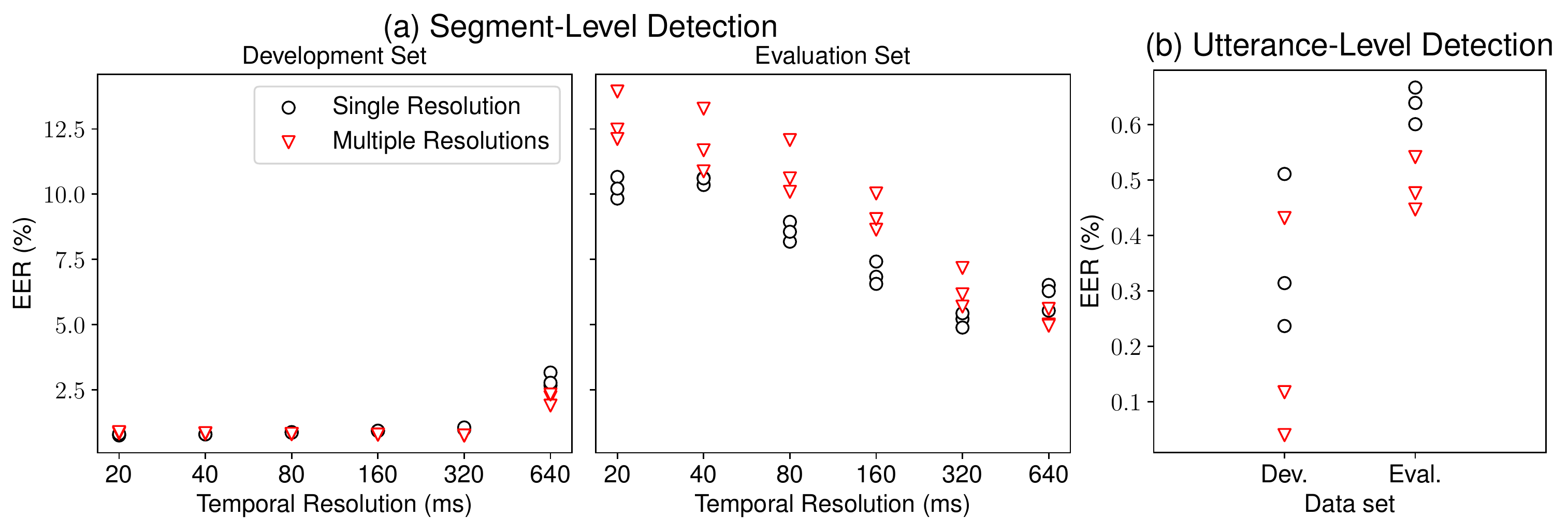}
\caption{{Comparison of proposed CMs trained at single temporal resolution and multiple temporal resolutions training strategies. Two figures on the left side show EERs of segment-level detection, and the right-side figure shows EERs of utterance-level detection. Each CM was trained three times with different random seeds.}}
\label{fig:eer-seg-utt}
\vspace{-4mm}
\end{figure*}

In this section, we compared {the two} CM training strategies, {that is, training at a single temporal resolution or multiple resolutions}. Performance was measured on the development set {and evaluation set} of the PartialSpoof database.

{We first explored the two training strategies on segment-level detection. The results are shown in Figure~\ref{fig:eer-seg-utt}(a). For the proposed CM trained at a single temporal resolution (black circle),} since the segment-level ground-truth labels are available at the six segment-level temporal resolutions, we trained six versions of the proposed CM, one for each segment-level temporal resolution. After training, each CM version produces scores at the corresponding temporal resolution. {We also trained the CM at multiple temporal resolutions (red triangle) that can derive scores for all resolutions at once.}

Results suggest that segment-level detection is more difficult at a higher temporal resolution, but it is not impossible. For the highest temporal resolution of 20 ms, the proposed CM achieved an EER of around 10\% on the evaluation set. This EER is reasonably good considering the fact that each segment consists of only one frame and is extremely short. {Besides, although the multiple-resolution CM is slightly better than the single-resolution counterpart at the resolution of 640 ms, it performed worse than the single-resolution CMs at fine-grained segment levels (20 $\sim$ 320 ms)}. There is thus room for improvement of the fine-grained segment-level detection.

{We then compared performance on utterance-level detection. The results are shown in Figure~\ref{fig:eer-seg-utt}(b). No matter which training strategy was used, the} proposed CM achieved an EER below 1\% on the evaluation set for utterance level detection. {Furthermore, the CM trained at multiple temporal resolutions} outperformed the CMs trained at a single temporal resolution for utterance-level detection. This indicates that the coarse-grained detection tasks can benefit from the training strategies using multiple temporal resolutions. 

Results of the above experiments suggest that we need to select a suitable temporal resolution and training strategy depending on our goal: 
{(1) For the} detection at the segment level, training should be applied at the specific target temporal resolution; 
{(2) in terms of} utterance-level detection, the use of more fine-grained information is {expected to improve CM performance.}

{Besides, the differences} between the EERs on the evaluation and development sets demonstrate the difficulty of {segment-level} detection and show that it is difficult to generalize to the evaluation set. {We thus explore the differences between them in the next subsection.}

\subsection{{EER gap between development and evaluation sets}}

\begin{figure}[!t]
\centering
\includegraphics[width=0.9\linewidth]{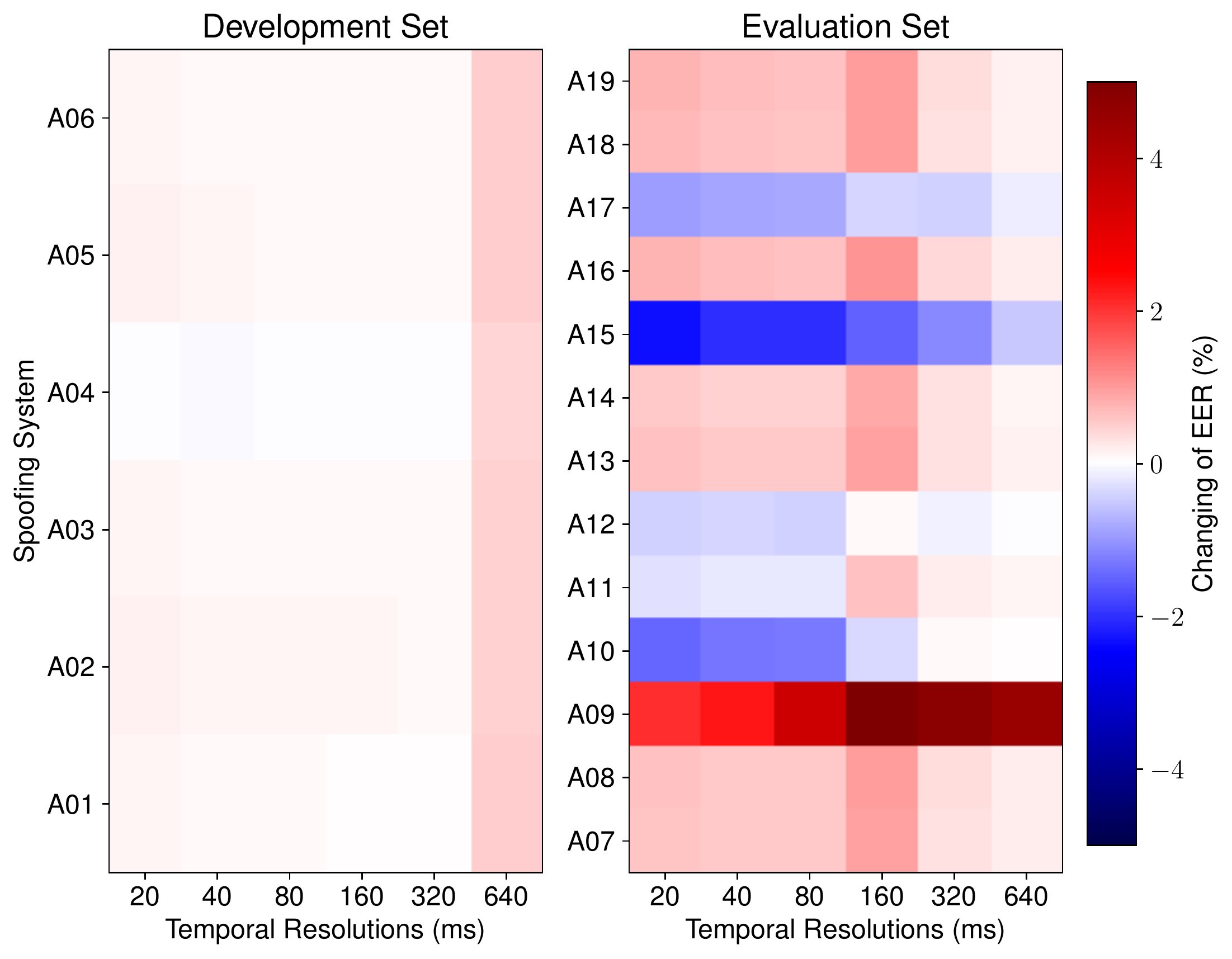}
\caption{{Results of leave-one-out experiments on development and evaluation sets. Color represents changing of EER (\%). }}
\label{fig:leave-one-out}
\vspace{-5mm}
\end{figure}

{
We have two hypotheses to explain the significant differences between the EERs on the development and evaluation sets, and we examined both using the CM trained at multiple temporal resolutions.}

{
(1) \textbf{Hypothesis 1}: \textit{More difficult spoofing systems exist in the evaluation set.} For the LA scenario and ASVspoof 2019 LA database, existing studies have measured the EER over fully spoofed trials from each spoofing system and found that a voice-conversion-based spoofing system called A17 significantly increased the CMs' EERs on the evaluation set \cite{Nautsch2021spoof19, tak2020end}. Since the PartialSpoof database used spoofed trials from the ASVspoof 2019 LA database, we hypothesize that the strong attacks in the evaluation set may have led to the higher EER. Because the segments in a partially-spoofed audio sample can be from different spoofing methods, we did not use the same investigation method as previous studies on the LA scenario and instead used a leave-one-out evaluation approach. For each spoofing method, we excluded the test trials that contain at least one segment produced by that spoofing method. We then measured the EER on the remaining trials and compared it with the original EER computed on all the test trials. We conducted the analysis on both the development and evaluation sets. 
} 

{The results are shown in Figure~\ref{fig:leave-one-out}. Each column in the figure corresponds to the results on segment-level detection at one temporal resolution, while each row corresponds to the results of analysis on a specific spoofing method. The IDs of the spoofing systems such as A07 were inherited from those in the ASVspoof 2019 LA database. It was observed that, if the spoofed trials that contain spoofed segments from A15\footnote{{A15 is a hybrid spoofing system that uses a TTS voice as a source speaker and WaveNet as a vocoder for voice conversion.}} were removed, the EER on the rest of the evaluation set decreased significantly. For our CM, A15 was the strongest spoofing system in the PS scenario. Leaving out other spoofing systems such as A10, A11, A12 and A17 also decreased the EER but to a lesser extent. In contrast, the EERs on the development set did not decrease as much as the change in the EERs on the evaluation set no matter which spoofing method was left out.
} 

{As hypothesized, partially-spoofed segments produced by some of the spoofing systems in the evaluation set were more difficult to detect. Particularly, A15 was the strongest attack. This may be one reason that EERs in the evaluation set are higher than those in the development set.
However, the EER on the evaluation set was still higher than on the development set even when we removed A15. We thus have another hypothesis to explain the EER gap between the development and evaluation sets.
}

\begin{figure}[!t]
\centering
\includegraphics[width=1\linewidth]{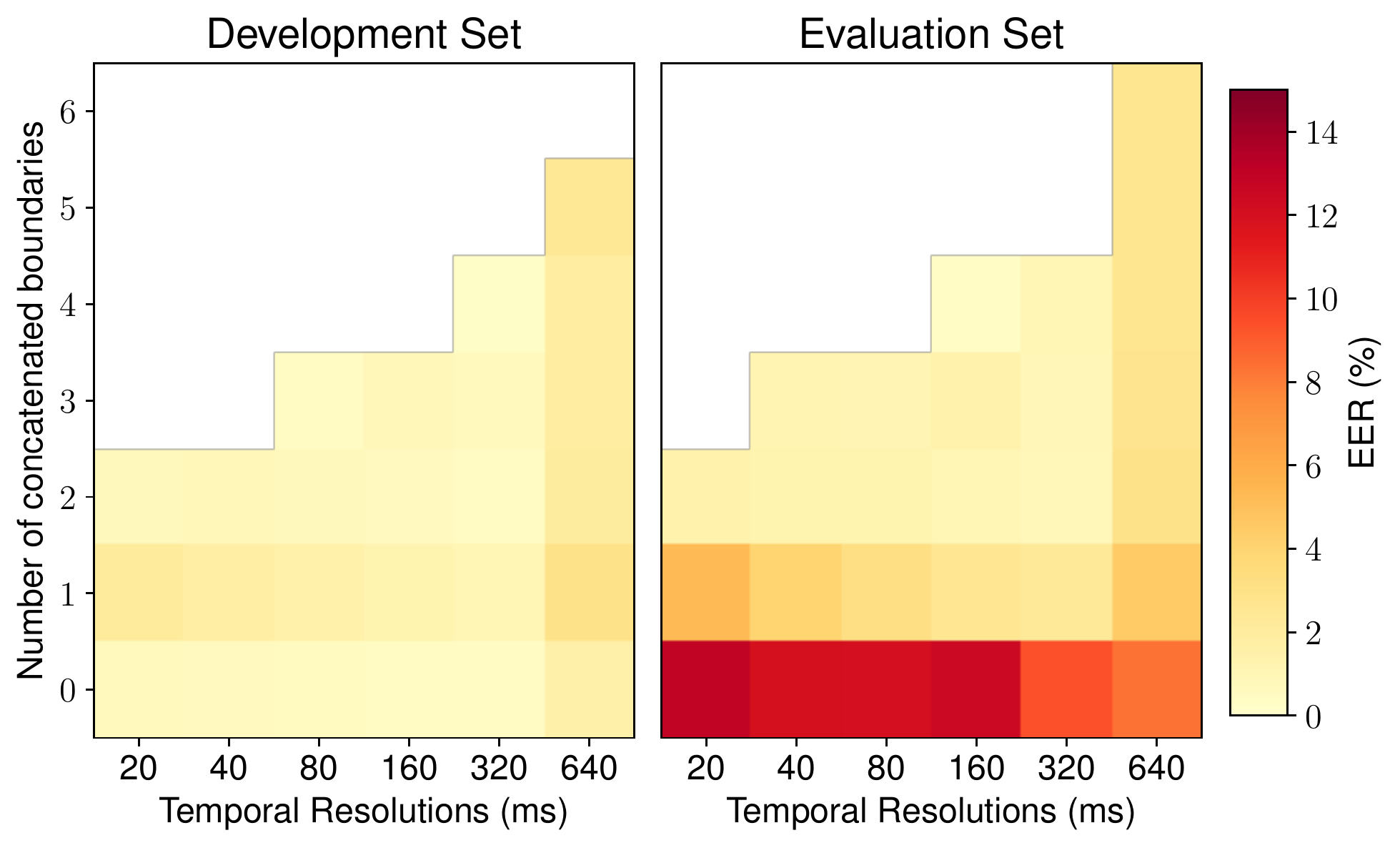}
\caption{{Breakdown of EER on number of concatenated boundaries.}}
\label{fig:joint}
\vspace{-5mm}
\end{figure}

{ 
(2) \textbf{Hypothesis 2}: \textit{Detecting spoofed segments with fewer concatenated boundaries on the evaluation set is more difficult than on the development set.}} 
{As described in Section~\ref{sec:database-procedures}, the partially-spoofed trials in PartialSpoof were created on the basis of substitution and concatenation. The overlap-add-based concatenation may bring in artifacts around the concatenated boundaries, which are expected to be useful for the CM. However, a segment to be scored by the CM may not contain any concatenated boundaries. Furthermore, if such a segment is from an unseen spoofing attack, the CM is more likely to make a mistake. Hence, we hypothesize that our CM performed worse on the evaluation set because it was not perfectly designed to distinguish the spoofed segments without any concatenation point from those with one or more concatenated boundaries. The performance on the development set may receive less impact since all the spoofing methods were seen during the training of the CM.
}

{ 
To verify the hypothesis, we computed the breakdown EERs on the basis of the number of concatenated boundaries in the segments. The results are listed in Figure~\ref{fig:joint}. Each row in the figure lists the EERs evaluated on segments that have a corresponding number of concatenated boundaries, and each column is for EERs at one temporal resolution. We observe that the performance of our CM on the evaluation set degraded when the number of concatenated boundaries decreased. Especially, when the number of concatenated boundaries was zero, the CM's EER significantly increased at temporal resolutions ranging from 20 to 160 ms. In contrast, the EERs on the development set varied (but to a much lesser extent) regardless of the change in the number of concatenated boundaries. As hypothesized, our CM made more errors when facing unseen spoofed segments with few concatenated boundaries in the evaluation set.
}

\begin{table*}[!t]
\begin{threeparttable}[b]
\centering
\footnotesize  
\caption{EERs (\%) of different CMs on PartialSpoof evaluation set. Column Types lists each CM in accordance with categories in Figure~\ref{fig:fully_CM}. All CMs were trained on PartialSpoof training set, although they may have used labels at different temporal resolutions. Previous studies \cite{Zhang2021PartialSpoof, zhang21partialspoof_mtl, Yi2021halftruth} used only one or two temporal resolutions.}
\label{tab:allres_partialspoof_temporary}
\begin{tabular}{ccccccccccccc}
\toprule

                                 & Model ID\tnote{1}      & Types      & Training          & Front-         & Back-                & \multicolumn{7}{c}{Temporal resolutions (ms)}                                                         \\
\cmidrule(l){7-13} 
Ref.                            & {(in ref.)} & {(in Fig.~\ref{fig:fully_CM})} & Resolutions & end & end & {20} & {40} & {80} & {160} & {320} & {640} & {utt.} \\
\midrule
\multirow{2}{*}{{\cite{Yi2021halftruth}}}  & {Utterance}    & {a}      & {utt.}             & \multirow{2}{*}{{CQCC}}               & \multirow{2}{*}{{LCNN}}                        &   &   &   &     &    &    & {17.77}          \\
  & {Segment\tnote{2}} & {c} & {10 ms } & & & {27.17} & & &    &    &    &           \\

\midrule
\multirow{2}{*}{\cite{Zhang2021PartialSpoof}}  & Utterance    &a      & utt.             & \multirow{2}{*}{LFCC\tnote{3}}               & \multirow{2}{*}{LCNN-BLSTM}                        &   &   &   & 40.20     &    &    & 6.19          \\
                                              & Segment  & c    & 160 ms             &                &                         &   &   &   & 16.21        &    &    & 8.61          \\
\midrule
\multirow{7}{*}{\cite{zhang21partialspoof_mtl}}           & Utterance & d     & utt.          & \multirow{7}{*}{LFCC}               & \multirow{7}{*}{SELCNN-BLSTM}                    &   &   &   & 44.00       &    &    & 6.33          \\
                                               & Segment     & c    & 160 ms           &                &                    &   &   &   & 15.93        &    &    & 7.69          \\
                                               & UttU        & d    & 160 ms, utt.     &                &                    &   &   &   & 20.04        &    &    & 9.96          \\
                                               & SegU        & c    & 160 ms, utt.     &                &                    &   &   &   & 17.75        &    &    & 7.04          \\
                                             & MulBS         & b     & 160 ms, utt.     &                &                    &   &   &   & 17.55        &    &    & 5.90           \\
                                             & UttBW         & b     & 160 ms, utt.     &                &                    &   &   &   & 17.77        &    &    & 5.66          \\
                                             & SegBW         & b     & 160 ms, utt.     &                &                    &   &   &   & 16.60         &    &    & 6.07          \\
\midrule
 {Proposed}              &  \texttt{{Multi reso.}}    & {e}   & 20$\sim$640, utt.          & w2v2-large         & 5gmlp                  & 12.84       & 11.94       & 10.92       & \phantom{0}\textbf{9.24}         & 6.34         & 5.19         & \textbf{0.49}          \\
\bottomrule
\end{tabular}
     \begin{tablenotes}
       \item [1] {The Model ID in the second column is defined by the corresponding references in the first column Ref.}
       \item [2] {The original resolution in \cite{Yi2021halftruth} was 8 ms, a value determined by the Constant Q Cepstral Coefficients (CQCC) extraction configuration \cite{todisco2017cqcc}. We used 10 ms so that the results from this system are comparable to those of the other systems. Results for 20 ms were calculated using the average method \cite{Yi2021halftruth}.}
       \item [3] LFCC stands for linear frequency cepstral coefficients.
     \end{tablenotes}
    \end{threeparttable}
\end{table*}

\subsection{Comparing proposed and conventional CMs}

We next compared the proposed CM with the {conventional} CMs  {\cite{Yi2021halftruth,Zhang2021PartialSpoof,zhang21partialspoof_mtl}} explained in Section~\ref{sec:previous-cms}. 
Comparisons are made to a CM trained with the {multiple temporal resolutions}. EERs for the PartialSpoof evaluation set are summarized in Table~\ref{tab:allres_partialspoof_temporary}.

The first noteworthy observation was that the conventional CMs could only {directly} detect segments at a single temporal resolution. This is due to the inflexible back-end structure, as discussed in Section~\ref{sec:overview_CM_remark}. In contrast, our proposed CM can detect spoofed segments at different temporal resolutions {simultaneously}. It also significantly outperformed the conventional CMs at the corresponding segment and utterance levels. These results indicate that the proposed CM is more suitable for the PS scenario.

\subsection{Cross-scenario study}

Since a CM for the PS scenario can execute utterance-level detection, it can also be applied to the LA scenario. To determine whether the improvements brought by multi-resolution training for the PS scenario also translate to the LA scenario, we investigated the performance of the proposed CM on utterance-level detection for both LA and PS scenarios. The data for these scenarios were from the ASVspoof 2019 LA and PartialSpoof databases, respectively.

These experiments involved two training settings, one using the training set of the ASVspoof 2019 LA database and the other using that of the PartialSpoof database. For the ASVspoof 2019 LA database, since the training data only contains utterance-level labels, the proposed CM was trained using the single temporal resolution training strategy for utterance-level detection. For CMs trained using the PartialSpoof data, we used the {two CMs trained using either the single temporal resolution strategy or the multiple resolution one as} described in Section \ref{sec:proposed-training-strategies}. The first used only utterance-level labels and was trained using the same strategy as the CM trained on the ASVspoof 2019 LA database. The {latter} used the full set of utterance and segment labels from the PartialSpoof database and trained using the multiple temporal resolution strategy. 

We used each CM to produce scores for the ASVspoof 2019 LA and PartialSpoof development and evaluation sets for the utterance-level detection. The results are shown in Table~\ref{tab:cross-database}. Focusing first on single resolution and utterance-level training, the performance for the ASVspoof 2019 LA database was shown to be competitive. However, the EERs increased significantly beyond 10\% in the PartialSpoof dataset. This was expected because the CMs trained using the LA data are not exposed to partial spoofs, thus perform poorly. When trained using the PartialSpoof dataset; however, performance in the case of PS data improved significantly, while performance when tested using the LA data remained competitive.

Further improvements were achieved when the proposed CM was trained at multiple temporal resolutions. In summary, when trained on the PartialSpoof dataset, the proposed CM demonstrated better utterance-level detection performance for both LA and PS scenarios. Using PartialSpoof training data {is} beneficial even when the CM is tested on LA data.\footnote{{In addition to the cross-scenario study, it is also possible to combine the two databases and see if they are complementary to each other. This investigation requires significant changes in the proposed CM structure so that model training can be effectively conducted even in situations where parts of audio files in the training database do not have segment-level labels, which is beyond the scope of this paper and is a future analysis topic.}}

\begin{table}[!t]
\centering
\footnotesize  
\caption{Cross-scenario study in utterance level (EER \%). (ASVspoof 2019 LA and PartialSpoof databases were applied for LA and PS scenarios separately.)} 
\label{tab:cross-database}
\setlength{\tabcolsep}{4pt}
\begin{tabular}{c|c|cc|cc}
\toprule
        &       & \multicolumn{2}{c|}{{LA}}  & \multicolumn{2}{c}{{PS}}     \\
&{Train} & {Dev.} & {Eval.}          & {Dev.} & {Eval.}  \\
\midrule
Single reso. at utt. level       & \multicolumn{1}{c|}{LA}   & 0.04      & 0.83     & 12.22 & 14.19 \\
\midrule
Single reso. at utt. level       & \multirow{2}{*}{PS} & 0.57      & 0.77      & \phantom{0}0.35  & \phantom{0}0.64  \\
Multiple reso. at all levels   &                               & 0.24      & 0.90      & \phantom{0}0.20  & \phantom{0}0.49  \\
\bottomrule
\end{tabular}
\vspace{-4mm}
\end{table}

\section{Conclusion}
\label{sec:conclusion}
We reported a new spoofing scenario, PS, in which only a fraction of speech segments are spoofed, with the remaining segments containing bona fide speech. Successful approaches to spoofing detection in this scenario can be applied at either the utterance level or segment level. The latter requires the assessment of spoofing classifiers using a database of bona fide and spoofed speech, with the latter labeled at the segment level. We described the new PartialSpoof database which is labeled at multiple temporal resolutions from 20 to 640 ms.

After formulating CM tasks required to tackle the PS scenario, we introduced SSL models as an enhanced front-end and proposed new neural architectures and training strategies that exploit segment-level labels for simultaneous, multi-resolution training. Experimental results suggest that CMs and training strategies should be adapted to a specific goal. Utterance-level detection can benefit from the use of more fine-grained information during training, whereas the comparatively more challenging task of segment-level detection calls for matched resolution. 

{
There were significant differences between the results of segment-level detection on the development and evaluation sets. We thus investigated two hypotheses to explain the differences: (1) More difficult spoofing systems exist in the evaluation set. This hypothesis is supported by a series of leave-one-out experiments for each resolution, and significant changes in the EER were observed after removing certain unknown spoofing systems. Among those unknown spoofing systems, A15 was found to be the strongest attack in the PartialSpoof dataset. (2) Detecting spoofed segments with fewer concatenated boundaries on the evaluation set is more difficult than on the development set. This is supported by an analysis of the breakdown EERs on the spoofed data with different numbers of concatenated boundaries. The performance of the CM on the evaluation set was much worse when there were fewer concatenated boundaries within the segments. In contrast, the performance on the development set was less affected. How to overcome this issue for the segment-level detection is worth exploring in the future.
}

{
We also conducted a cross-scenario study on the LA and PS scenarios. }
The proposed CM was shown to achieve the best reported utterance-level detection result for the ASVspoof 2019 LA database (an EER of {0.77}\% for the evaluation set). The PS scenario is a realistic, important, timely, and challenging spoofing scenario, which warrants greater attention in the future. The more conventional, utterance-level CMs considered in {the LA scenario} largely fail in the face of only partially-spoofed utterances, indicating their vulnerability to manipulation through such attacks. Results nonetheless show scope for improvement. 

Our future work will include explicit use of linguistic information for the detection of short spoofed segments embedded in otherwise bona fide utterances {and robust training strategies to handle imbalanced classes.} {An important challenge will be to increase the number of TTS and VC methods for creating partially-spoofed speech and to use more up-to-date methods.}

\bibliographystyle{IEEEtran}
\bibliography{main}

\begin{thebibliography}{10}
\providecommand{\url}[1]{#1}
\csname url@samestyle\endcsname
\providecommand{\newblock}{\relax}
\providecommand{\bibinfo}[2]{#2}
\providecommand{\BIBentrySTDinterwordspacing}{\spaceskip=0pt\relax}
\providecommand{\BIBentryALTinterwordstretchfactor}{4}
\providecommand{\BIBentryALTinterwordspacing}{\spaceskip=\fontdimen2\font plus
\BIBentryALTinterwordstretchfactor\fontdimen3\font minus
  \fontdimen4\font\relax}
\providecommand{\BIBforeignlanguage}[2]{{%
\expandafter\ifx\csname l@#1\endcsname\relax
\typeout{** WARNING: IEEEtran.bst: No hyphenation pattern has been}%
\typeout{** loaded for the language `#1'. Using the pattern for}%
\typeout{** the default language instead.}%
\else
\language=\csname l@#1\endcsname
\fi
#2}}
\providecommand{\BIBdecl}{\relax}
\BIBdecl

\bibitem{Wu2014}
Z.~Wu, T.~Kinnunen, N.~Evans, J.~Yamagishi, C.~Hanil{\c{c}}i, M.~Sahidullah,
  and A.~Sizov, ``{ASVspoof 2015: the First Automatic Speaker Verification
  Spoofing and Countermeasures Challenge},'' in \emph{Proc. Interspeech}, 2015,
  pp. 2037--2041.

\bibitem{Kinnunen2017}
T.~Kinnunen, M.~Sahidullah, H.~Delgado, M.~Todisco, N.~Evans, J.~Yamagishi, and
  K.~A. Lee, ``{The ASVspoof 2017 Challenge: Assessing the Limits of Replay
  Spoofing Attack Detection},'' in \emph{Proc. Interspeech}, 2017, pp. 2--6.

\bibitem{Nautsch2021spoof19}
A.~Nautsch, X.~Wang, N.~Evans, T.~H. Kinnunen, V.~Vestman, M.~Todisco,
  H.~Delgado, M.~Sahidullah, J.~Yamagishi, and K.~A. Lee, ``Asvspoof 2019:
  Spoofing countermeasures for the detection of synthesized, converted and
  replayed speech,'' \emph{IEEE Transactions on Biometrics, Behavior, and
  Identity Science}, vol.~3, no.~2, pp. 252--265, 2021.

\bibitem{asvspoof2021}
J.~Yamagishi, X.~Wang, M.~Todisco, M.~Sahidullah, J.~Patino, A.~Nautsch,
  X.~Liu, K.~A. Lee, T.~Kinnunen, N.~Evans, and H.~Delgado, ``{ASVspoof 2021:
  accelerating progress in spoofed and deepfake speech detection},'' in
  \emph{Proc. ASVspoof 2021 Workshop}, 2021, pp. 47--54.

\bibitem{Zhang2021PartialSpoof}
L.~Zhang, X.~Wang, E.~Cooper, J.~Yamagishi, J.~Patino, and N.~Evans, ``{An
  Initial Investigation for Detecting Partially Spoofed Audio},'' in
  \emph{Proc. Interspeech}, 2021, pp. 4264--4268.

\bibitem{zhang21partialspoof_mtl}
L.~Zhang, X.~Wang, E.~Cooper, and J.~Yamagishi, ``{Multi-task Learning in
  Utterance-level and Segmental-level Spoof Detection},'' in \emph{Proc.
  ASVspoof 2021 Workshop}, 2021, pp. 9--15.

\bibitem{BaevskiZMA20-w2v2}
A.~Baevski, Y.~Zhou, A.~Mohamed, and M.~Auli, ``wav2vec 2.0: A framework for
  self-supervised learning of speech representations,'' in \emph{Proc.
  NeurIPS}, vol.~33, 2020, pp. 12\,449--12\,460.

\bibitem{chen2021wavlm}
S.~Chen, C.~Wang, Z.~Chen, Y.~Wu, S.~Liu, Z.~Chen, J.~Li, N.~Kanda,
  T.~Yoshioka, X.~Xiao, J.~Wu, L.~Zhou, S.~Ren, Y.~Qian, Y.~Qian, J.~Wu,
  M.~Zeng, X.~Yu, and F.~Wei, ``Wavlm: Large-scale self-supervised pre-training
  for full stack speech processing,'' \emph{IEEE Journal of Selected Topics in
  Signal Processing}, vol.~16, no.~6, pp. 1505--1518, 2022.

\bibitem{Zakariah2018-forgery}
M.~Zakariah, M.~K. Khan, and H.~Malik, ``Digital multimedia audio forensics:
  past, present and future,'' \emph{Multimedia Tools and Applications},
  vol.~77, pp. 1009--1040, 2018.

\bibitem{bevinamarad2020audioforgery}
P.~R. Bevinamarad and M.~Shirldonkar, ``Audio forgery detection techniques:
  Present and past review,'' in \emph{Proc. ICOEI}.\hskip 1em plus 0.5em minus
  0.4em\relax IEEE, 2020, pp. 613--618.

\bibitem{campbell1996chatr}
N.~Campbell, ``Chatr: A high-definition speech re-sequencing system,'' in
  \emph{Proc. 3rd ASA/ASJ Joint meeting}, 1996, pp. 1223--1228.

\bibitem{PaulTaylor}
P.~Taylor, \emph{{Text-to-Speech Synthesis}}.\hskip 1em plus 0.5em minus
  0.4em\relax Cambridge University Press, 2009.

\bibitem{Tan_2021_editspeech}
D.~Tan, L.~Deng, Y.~T. Yeung, X.~Jiang, X.~Chen, and T.~Lee, ``Editspeech: A
  text based speech editing system using partial inference and bidirectional
  fusion,'' in \emph{Proc. ASRU}, 2021, pp. 626--633.

\bibitem{morrison2021context}
M.~Morrison, L.~Rencker, Z.~Jin, N.~J. Bryan, J.-P. Caceres, and B.~Pardo,
  ``Context-aware prosody correction for text-based speech editing,'' in
  \emph{Proc. ICASSP}, 2021, pp. 7038--7042.

\bibitem{Descript}
\BIBentryALTinterwordspacing
``{Descript}.'' [Online]. Available: \url{https://www.descript.com/}
\BIBentrySTDinterwordspacing

\bibitem{Yi2021halftruth}
J.~Yi, Y.~Bai, J.~Tao, H.~Ma, Z.~Tian, C.~Wang, T.~Wang, and R.~Fu,
  ``{Half-Truth: A Partially Fake Audio Detection Dataset},'' in \emph{Proc.
  Interspeech}, 2021, pp. 1654--1658.

\bibitem{Yi2022ADD}
J.~Yi, R.~Fu, J.~Tao, S.~Nie, H.~Ma, C.~Wang, T.~Wang, Z.~Tian, Y.~Bai, C.~Fan,
  S.~Liang, S.~Wang, S.~Zhang, X.~Yan, L.~Xu, Z.~Wen, and H.~Li, ``Add 2022:
  the first audio deep synthesis detection challenge,'' in \emph{Proc. ICASSP},
  2022, pp. 9216--9220.

\bibitem{Wang2020data}
X.~Wang, J.~Yamagishi, M.~Todisco, H.~Delgado, A.~Nautsch, N.~Evans,
  M.~Sahidullah, V.~Vestman, T.~Kinnunen, K.~A. Lee, L.~Juvela, P.~Alku, Y.-H.
  Peng, H.-T. Hwang, Y.~Tsao, H.-M. Wang, S.~L. Maguer, M.~Becker,
  F.~Henderson, R.~Clark, Y.~Zhang, Q.~Wang, Y.~Jia, K.~Onuma, K.~Mushika,
  T.~Kaneda, Y.~Jiang, L.-J. Liu, Y.-C. Wu, W.-C. Huang, T.~Toda, K.~Tanaka,
  H.~Kameoka, I.~Steiner, D.~Matrouf, J.-F. Bonastre, A.~Govender, S.~Ronanki,
  J.-X. Zhang, and Z.-H. Ling, ``Asvspoof 2019: A large-scale public database
  of synthesized, converted and replayed speech,'' \emph{Computer Speech and
  Language}, vol.~64, p. 101114, 2020.

\bibitem{sv56}
{International Telecommunication Union}, Recommendation {G.191}: Software Tools
  and Audio Coding Standardization, Nov 11 2005.

\bibitem{povey2011kaldi}
D.~Povey, A.~Ghoshal, G.~Boulianne, L.~Burget, O.~Glembek, N.~Goel,
  M.~Hannemann, P.~Motlicek, Y.~Qian, P.~Schwarz \emph{et~al.}, ``The {Kaldi}
  speech recognition toolkit,'' in \emph{Proc. ASRU}, 2011.

\bibitem{kinnunen2010overview}
T.~Kinnunen and H.~Li, ``{An overview of text-independent speaker recognition:
  From features to supervectors},'' \emph{Speech communication}, vol.~52,
  no.~1, pp. 12--40, 2010.

\bibitem{Lavechin-sad-dihard}
M.~Lavechin, M.-P. Gill, R.~Bousbib, H.~Bredin, and L.~{Paola Garcia-Perera},
  ``{End-to-end Domain-Adversarial Voice Activity Detection},'' in \emph{Proc.
  Interspeech}, 2020, pp. 3685--3689.

\bibitem{martin2022ADD1}
J.~M. Martín-Doñas and A.~Álvarez, ``The vicomtech audio deepfake detection
  system based on wav2vec2 for the 2022 add challenge,'' in \emph{Proc.
  ICASSP}, 2022, pp. 9241--9245.

\bibitem{haibin2022partialQA}
H.~Wu, H.-C. Kuo, N.~Zheng, K.-H. Hung, H.-Y. Lee, Y.~Tsao, H.-M. Wang, and
  H.~Meng, ``Partially fake audio detection by self-attention-based fake span
  discovery,'' in \emph{Proc. ICASSP}, 2022, pp. 9236--9240.

\bibitem{yang21c_interspeech_superb}
S.~wen Yang, P.-H. Chi, Y.-S. Chuang, C.-I.~J. Lai, K.~Lakhotia, Y.~Y. Lin,
  A.~T. Liu, J.~Shi, X.~Chang, G.-T. Lin, T.-H. Huang, W.-C. Tseng, K.~tik Lee,
  D.-R. Liu, Z.~Huang, S.~Dong, S.-W. Li, S.~Watanabe, A.~Mohamed, and
  H.~yi~Lee, ``{SUPERB: Speech Processing Universal PERformance Benchmark},''
  in \emph{Proc. Interspeech}, 2021, pp. 1194--1198.

\bibitem{wang2021ssl}
X.~Wang and J.~Yamagishi, ``{Investigating Self-Supervised Front Ends for
  Speech Spoofing Countermeasures},'' in \emph{Proc. Odyssey}, 2022, pp.
  100--106.

\bibitem{JoelShorSSL2021}
J.~Shor, A.~Jansen, W.~Han, D.~Park, and Y.~Zhang, ``Universal paralinguistic
  speech representations using self-supervised conformers,'' in \emph{Proc.
  ICASSP}, 2022, pp. 3169--3173.

\bibitem{hsu2021hubert}
W.-N. Hsu, B.~Bolte, Y.-H.~H. Tsai, K.~Lakhotia, R.~Salakhutdinov, and
  A.~Mohamed, ``{HuBERT: Self-Supervised Speech Representation Learning by
  Masked Prediction of Hidden Units},'' \emph{IEEE/ACM Transactions on Audio,
  Speech, and Language Processing}, vol.~29, pp. 3451--3460, 2021.

\bibitem{vaswani2017attention}
A.~Vaswani, N.~Shazeer, N.~Parmar, J.~Uszkoreit, L.~Jones, A.~N. Gomez,
  {\L}.~Kaiser, and I.~Polosukhin, ``{Attention is all you need},'' in
  \emph{Proc. NeurIPS}, 2017, pp. 5998--6008.

\bibitem{wang2021comparative}
X.~Wang and J.~Yamagishi, ``{A Comparative Study on Recent Neural Spoofing
  Countermeasures for Synthetic Speech Detection},'' in \emph{Proc.
  Interspeech}, 2021, pp. 4259--4263.

\bibitem{godfrey1992switchboard}
J.~J. Godfrey, E.~C. Holliman, and J.~McDaniel, ``{SWITCHBOARD: Telephone
  speech corpus for research and development},'' in \emph{Proc. ICASSP}, 1992,
  pp. 517--520.

\bibitem{kahn2020libri}
J.~Kahn, M.~Rivi{\`{e}}re, W.~Zheng, E.~Kharitonov, Q.~Xu, P.-E. Mazar{\'{e}},
  J.~Karadayi, V.~Liptchinsky, R.~Collobert, C.~Fuegen, and Others,
  ``{Libri-light: A benchmark for ASR with limited or no supervision},'' in
  \emph{Proc. ICASSP}, 2020, pp. 7669--7673.

\bibitem{cieri-etal-2004-fisher}
C.~Cieri, D.~Miller, and K.~Walker, ``{The Fisher Corpus: a Resource for the
  Next Generations of Speech-to-Text},'' in \emph{Proc. LREC}, Lisbon,
  Portugal, May 2004.

\bibitem{panayotov2015librispeech}
V.~Panayotov, G.~Chen, D.~Povey, and S.~Khudanpur, ``{Librispeech: an ASRcorpus
  based on public domain audio books},'' in \emph{Proc. ICASSP}, 2015, pp.
  5206--5210.

\bibitem{ardila-etal-2020-common}
R.~Ardila, M.~Branson, K.~Davis, M.~Kohler, J.~Meyer, M.~Henretty, R.~Morais,
  L.~Saunders, F.~Tyers, and G.~Weber, ``{Common Voice: A
  Massively-Multilingual Speech Corpus},'' in \emph{Proc. LREC}, 2020, pp.
  4218--4222.

\bibitem{harper2011iarpa}
IARPA, ``{Babel program},'' 2011.

\bibitem{chen21o_GigaSpeech}
G.~Chen, S.~Chai, G.-B. Wang, J.~Du, W.-Q. Zhang, C.~Weng, D.~Su, D.~Povey,
  J.~Trmal, J.~Zhang, M.~Jin, S.~Khudanpur, S.~Watanabe, S.~Zhao, W.~Zou,
  X.~Li, X.~Yao, Y.~Wang, Z.~You, and Z.~Yan, ``{GigaSpeech: An Evolving,
  Multi-Domain ASR Corpus with 10,000 Hours of Transcribed Audio},'' in
  \emph{Proc. Interspeech}, 2021, pp. 3670--3674.

\bibitem{wang-etal-2021-voxpopuli}
C.~Wang, M.~Riviere, A.~Lee, A.~Wu, C.~Talnikar, D.~Haziza, M.~Williamson,
  J.~Pino, and E.~Dupoux, ``{V}ox{P}opuli: A large-scale multilingual speech
  corpus for representation learning, semi-supervised learning and
  interpretation,'' in \emph{Proc. ACL-IJCNLP}, Aug. 2021, pp. 993--1003.

\bibitem{hochreiter1997long}
S.~Hochreiter and J.~Schmidhuber, ``{Long short-term memory},'' \emph{Neural
  computation}, vol.~9, no.~8, pp. 1735--1780, 1997.

\bibitem{Liu2021gmlp}
H.~Liu, Z.~Dai, D.~So, and Q.~V. Le, ``Pay attention to mlps,'' in \emph{Proc.
  NeurIPS}, vol.~34.\hskip 1em plus 0.5em minus 0.4em\relax Curran Associates,
  Inc., 2021, pp. 9204--9215.

\bibitem{tak2020end}
H.~Tak, J.~Patino, M.~Todisco, A.~Nautsch, N.~Evans, and A.~Larcher,
  ``End-to-end anti-spoofing with rawnet2,'' in \emph{Proc. ICASSP}, 2021, pp.
  6369--6373.

\bibitem{todisco2017cqcc}
M.~Todisco, H.~Delgado, and N.~Evans, ``Constant q cepstral coefficients: A
  spoofing countermeasure for automatic speaker verification,'' \emph{Computer
  Speech \& Language}, vol.~45, pp. 516--535, 2017.

\bibitem{bengio2004statistical}
S.~Bengio and J.~Mari{\'{e}}thoz, ``{A statistical significance test for person
  authentication},'' in \emph{Proc. Odyssey}, 2004.

\end{thebibliography}

\balance
\begin{IEEEbiography}[{\includegraphics[width=1in,height=1.25in,clip,keepaspectratio]{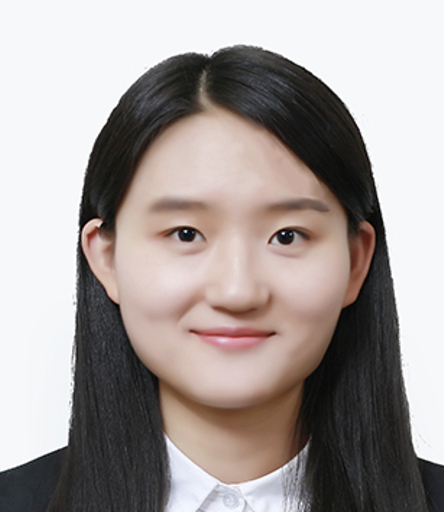}}]{Lin Zhang} (Student Member, IEEE) received an M.S. degree from Tianjin University, China, in 2020. Then, she worked as a Research Assistant with the SMIIP Lab at Duke Kunshan University in 2020. She is currently working toward the Ph.D. degree with the SOKENDAI/National Institute of Informatics, Japan. Her research interests include anti-spoofing, speaker recognition, and machine learning.
\end{IEEEbiography}

\begin{IEEEbiography}[{\includegraphics[width=1in,height=1.25in,clip,keepaspectratio]{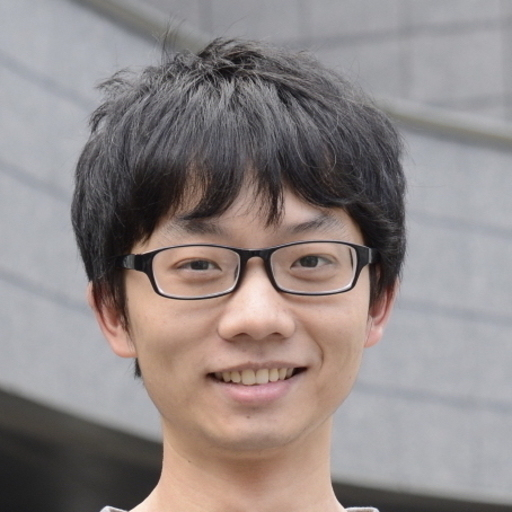}}]{Xin Wang} (Member, IEEE) is a project assistant professor at the National Institute of Informatics (NII), Japan. He received the Ph.D. degree from SOKENDAI/NII, Japan, in 2018. Before that, he received M.S. and B.E degrees from the University of Science and Technology of China and University of Electronic Science and Technology of China in 2015 and 2012, respectively. His research interests include statistical speech synthesis, speech security, and machine learning. He is a co-organizer of the latest ASVspoof and VoicePrivacy challenges. 
\end{IEEEbiography}

\begin{IEEEbiography}[{\includegraphics[width=1in,height=1.25in,clip,keepaspectratio]{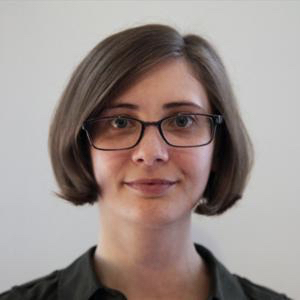}}]{Erica Cooper} (Member, IEEE) received a B.Sc. degree and M.Eng. degree both in electrical engineering and computer science from the Massachusetts Institute of Technology, Cambridge, MA, USA, in 2009 and 2010, respectively. She received a Ph.D. degree in computer science from Columbia University, New York, NY, USA, in 2019. Since 2019, she has been a Project Researcher with the National Institute of Informatics, Chiyoda, Tokyo, Japan. Her research interests include statistical machine learning and speech synthesis. Dr. Cooper's awards include the 3rd Prize in the CSAW Voice Biometrics and Speech Synthesis Competition, the Computer Science Service Award from Columbia University, and the Best Poster Award in the Speech Processing Courses in Crete.
\end{IEEEbiography}

\begin{IEEEbiography}[{\includegraphics[width=1in,height=1.25in,clip,keepaspectratio]{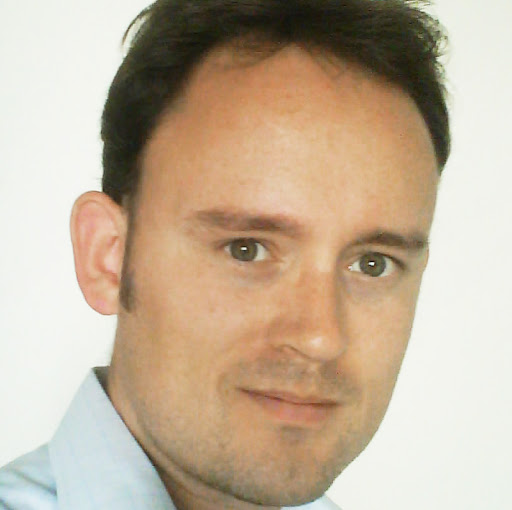}}]{Nicholas Evans} (Member, IEEE) is a Professor at EURECOM, France, where he heads research in Audio Security and Privacy. He is a co-founder of the community led, ASVspoof, SASV, and VoicePrivacy challenge series. He participated in the EU FP7 Tabula Rasa and EU H2020 OCTAVE projects, both involving antispoofing. Today, his team is leading the EU H2020 TReSPAsS-ETN project, a training initiative in security and privacy for multiple biometric characteristics. He co-edited the second and third editions of the Handbook of Biometric Anti-Spoofing, served previously on the IEEE Speech and Language Technical Committee and serves currently as an associate editor for the IEEE Trans. on Biometrics, Behavior, and Identity Science.
\end{IEEEbiography}

\begin{IEEEbiography}[{\includegraphics[width=1in,height=1.25in,clip,keepaspectratio]{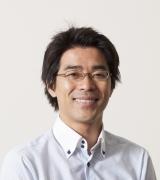}}]
{Junichi Yamagishi} (Senior Member, IEEE) received a Ph.D.\ degree from the Tokyo Institute of Technology (Tokyo Tech), Tokyo, Japan, in 2006. From 2007-2013, he was a research fellow in the Centre for Speech Technology Research (CSTR) at the University of Edinburgh, UK. He was appointed Associate Professor at the National Institute of Informatics, Japan, in 2013. He is currently a Professor at NII, Japan. His research topics include speech processing, machine learning, signal processing, biometrics, digital media cloning, and media forensics. 
He served previously as co-organizer for the bi-annual ASVspoof Challenge and the bi-annual Voice Conversion Challenge. He also served as a member of the IEEE Speech and Language Technical Committee (2013-2019), an Associate Editor of the IEEE/ACM Transactions on Audio Speech and Language Processing (2014-2017), and a chairperson of ISCA SynSIG (2017- 2021). He is currently a PI of JST-CREST and ANR supported VoicePersonae project and a Senior Area Editor of the IEEE/ACM TASLP.
\end{IEEEbiography}


\begin{appendices}
\section{Impact on ASV system}


This section demonstrates the threat to automatic speaker verification (ASV) systems in the PS scenario.
%
%
When segments generated by TTS and/or VC are embedded in an utterance, such a partially spoofed utterance may be misjudged as ``matched with enrolled users'' by ASV, especially when the speaker characteristics of the generated segments and the remaining part of the utterances are sufficiently similar, or when the percentage of the generated segments within the utterance is low. 

To empirically show this risk, we performed a pair of ASV experiments involving target trials,  zero-effort casual impostors and partially spoofed utterances. Both of the zero-effort casual impostors and partially spoofed utterances were treated as ``non-target'' trials. Number of target and non-target speakers were 10 and 10 respectively in the dev.\ set, and 48 and 19 in the eval.\ set. These experiments were performed using the same ASV system used for the ASVspoof 2019 challenge \cite{Wang2020data}. The front-end speaker embedding extractor is based on the Kaldi \cite{povey2011kaldi} pre-trained x-vector (m7) model.  The back-end is based on a probabilistic linear discriminant analysis (PLDA) model adapted using bona fide utterances contained in the train subset.


Results are presented in Table \ref{lab:ASV}. In the case of target utterances with only zero-effort impostors, the ASV system delivers equal error rates (EERs) of less than 3\% for both dev.\ and eval.\ sets. However, under the condition where partially spoofed audio samples are additionally introduced, the EERs increase substantially to unacceptably high levels.

\begin{table}[!htb]
\caption{Impact of partially spoofed audio on ASV system performance. Two types of non-target (zero-effort and partial spoof impostor) are considered. \label{lab:ASV}}
\setlength{\tabcolsep}{4pt}
\centering
\vspace{-3mm}
\begin{tabular}{ccc}
\toprule
\textbf{Conditions}        & \textbf{Dev. EER (\%)} & \textbf{Eval. EER (\%)} \\
\midrule
\textbf{Target + Zero-effort Impostors}       & \phantom{0}2.43        & \phantom{0}2.46              \\
\textbf{Target + Zero-effort + Partial Spoof} & 41.45       & 40.73            \\
\bottomrule
\end{tabular}
\end{table}

\begin{figure}[!b]
\centering
\includegraphics[width=.9\linewidth]{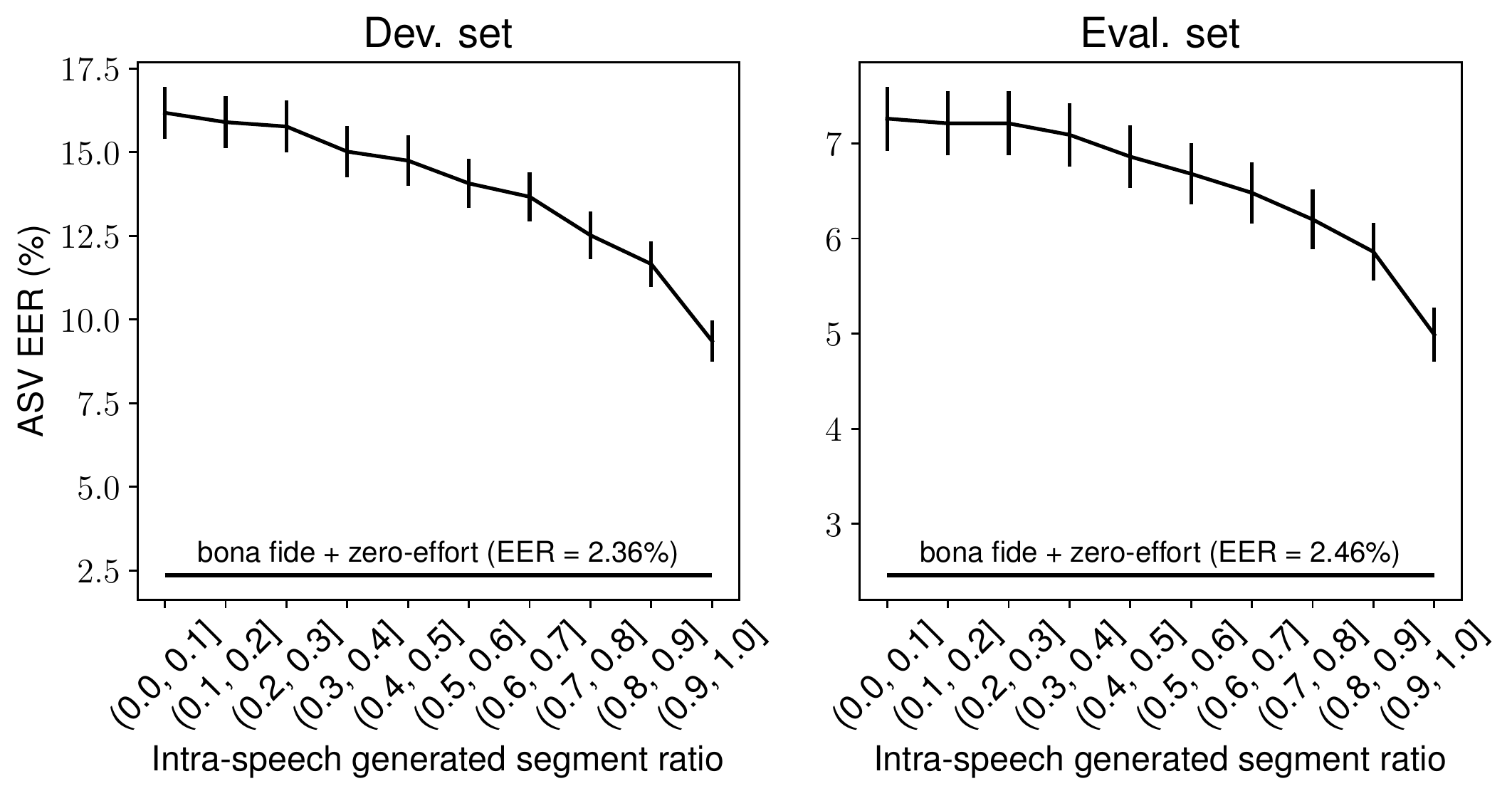}
\caption{ASV EERs for each of the quantized classes with the same number of trials, and confidence intervals at a significance level of 5\% \cite{bengio2004statistical}.}
\label{fig:asveer-ratio}
\end{figure}

To further analyze the impact on the ASV system, we divided the partially spoofed utterances into ten groups according to intra-speech generated segment ratios as introduced in Section~\ref{sec:database} and computed EERs of each group separately. Note that although the number of spoofed trials for each group is similar to each other, they are not identical. 
To make the EERs comparable to each other, we further randomly selected the same number of spoofed trials for each group. 
%
Figure~\ref{fig:asveer-ratio} shows the results. 
%
%
We can see that EERs increase inversely according to the intra-speech generated segment ratio as expected. 

%
%


\end{appendices}


\end{document}